\documentclass[a4paper,preprint2]{aastex}

\usepackage{natbib}
\usepackage{amsmath}
\usepackage{color, colortbl}
\definecolor{Gray}{gray}{0.9}

\textheight=\baselinestretch\textheight
\ifdim\textheight>25.2cm\textheight=25.0cm\fi

\topskip\baselineskip
\maxdepth\baselineskip
\let\tighten=\relax
\let\tightenlines=\tighten

\slugcomment{To appear in ApJ.}

\shorttitle{[GASS high velocity cloud catalogue}
\shortauthors{[V.A. Moss et al.}

\begin{document}

\title{Tracing dense and diffuse neutral hydrogen in the halo of the Milky Way} 

\author{
V~A.\,Moss\altaffilmark{1,2,3}$^{\dagger}$,
F~J.\,Lockman\altaffilmark{4},
and N~M.\,McClure-Griffiths\altaffilmark{5,3}
}

\altaffiltext{$\dagger$}{\bf vanessa.moss@sydney.edu.au}
\altaffiltext{1}{Sydney Institute for Astronomy, School of Physics A29, University of Sydney, NSW 2006, Australia}
\altaffiltext{2}{ARC Centre of Excellence for All-Sky Astrophysics (CAASTRO)}
\altaffiltext{3}{CSIRO Astronomy and Space Science, ATNF, PO Box 76, Epping, NSW 1710, Australia}
\altaffiltext{4}{National Radio Astronomy Observatory, PO Box 2, Green Bank, WV 24944}
\altaffiltext{5}{Research School of Astronomy and Astrophysics, Australian National University, Canberra, ACT 2611, Australia}

\begin{abstract}
We have combined observations of Galactic high-velocity H{\sc i} from two surveys: a very sensitive survey from the Green Bank 140\,ft Telescope with limited sky coverage, and the less sensitive but complete Galactic All Sky Survey from the 64\,m Parkes Radio Telescope. The two surveys preferentially detect different forms of neutral gas due to their sensitivity. We adopt a machine learning approach to divide our data into two populations that separate across a range in column density: 1) a narrow line-width population typical of the majority of bright high velocity cloud components, and 2) a fainter, broad line-width population that aligns well with that of the population found in the Green Bank survey. We refer to these populations as dense and diffuse gas respectively, and find that diffuse gas is typically located at the edges and in the tails of high velocity clouds, surrounding dense components in the core. A fit to the average spectrum of each type of gas in the Galactic All Sky Survey data reveals the dense population to have a typical line width of $\sim$20\,km~s$^{-1}$ and brightness temperature of $\sim$0.3\,K, while the diffuse population has a typical line width of $\sim$30\,km~s$^{-1}$ and a brightness temperature of $\sim$0.2\,K. Our results confirm that most surveys of high velocity gas in the Milky Way halo are missing the majority of the ubiquitous diffuse gas, and that this gas is likely to contribute at least as much mass as the dense gas.  
\end{abstract}



\section{Introduction}
The complicated galactic ecosystem of outflow, infall and circulation is known to play a key role in galaxy evolution. Outflowing gas in the form of supershells, chimneys and high-mass star formation provides an avenue for material to escape the disk into the haloes of galaxies \citep{McClureGriffiths:2006di,Ford:2010ga} potentially leading to thicker gaseous disks \citep{Oosterloo:2007kd,Kamphuis:2013hz} and metal transport to the halo \citep{Oey:2003wy,Werk:2011ht}. Conversely, infalling gas through mergers, interactions with other galaxies or interactions with the intergalactic medium drives change and causes disruption in galaxies and their environments \citep{Barnes:1992hb,Chung:2006bj,Westmeier:2013kq,Johnson:2013jp}, potentially providing new fuel for star formation \citep{Lockman:2008ha,Hill:2009gx,Tumlinson:2011by,Lehner:2011ic,Wolfe:2013cn}. In this paper, we aim to observationally uncover the hidden structure of faint neutral hydrogen in the halo of the Milky Way and how it relates to the prominent and well-known high-velocity cloud population.

Hints of a connection between the brighter HVCs and a faint diffuse background have been revealed in previous high-sensitivity work. \citet{Nigra:2012ds} studied a small isolated cloud in the Magellanic Stream and discovered an extremely faint broad component of width $\sim$60\,km~s$^{-1}$ co-located with the brighter 20\,km~s$^{-1}$ main component. They interpreted this multi-phase cloud as a warm neutral medium/warm ionised medium core surrounded by a turbulent mixing layer, giving insight into the potential connection between the neutral and ionised components of hydrogen in the halo. 

Studies of head-tail or core-envelope HVCs have revealed clear signatures of interaction between the clouds and their surrounding halo environment \citep[e.g.][]{Bruns:2000wt,Bruns:2001fw,BenBekhti:2006ef,For:2013is}, though with focus on individual clouds or subsets of clouds in a specific region rather than on the overall neutral hydrogen sky. On a large scale, \citet{Putman:2011gra} used H{\sc i} Parkes All Sky Survey (HIPASS) data to study the distribution of head-tail clouds across the southern sky, finding that roughly half could be attributed to the Magellanic system and overall these clouds appeared to have relatively short lifetimes. Moving away from clouds and towards overall halo structure, \citet{Murphy:1995hn} carried out a sensitive search for Galactic high-velocity H{\sc i} clouds towards quasars, finding most of their new detections to be associated with known HVC complexes and a large covering factor for Galactic HVCs. They also concluded that their observed faint high-velocity gas was low column density H{\sc i} that surrounded HVC complexes, rather than forming a completely separate population. This study later evolved into the sample of \citet{Lockman:2002bu}.

\begin{figure*}
  \centering
  \includegraphics[width=0.45\textwidth, angle=0, trim=0 0 0 0]{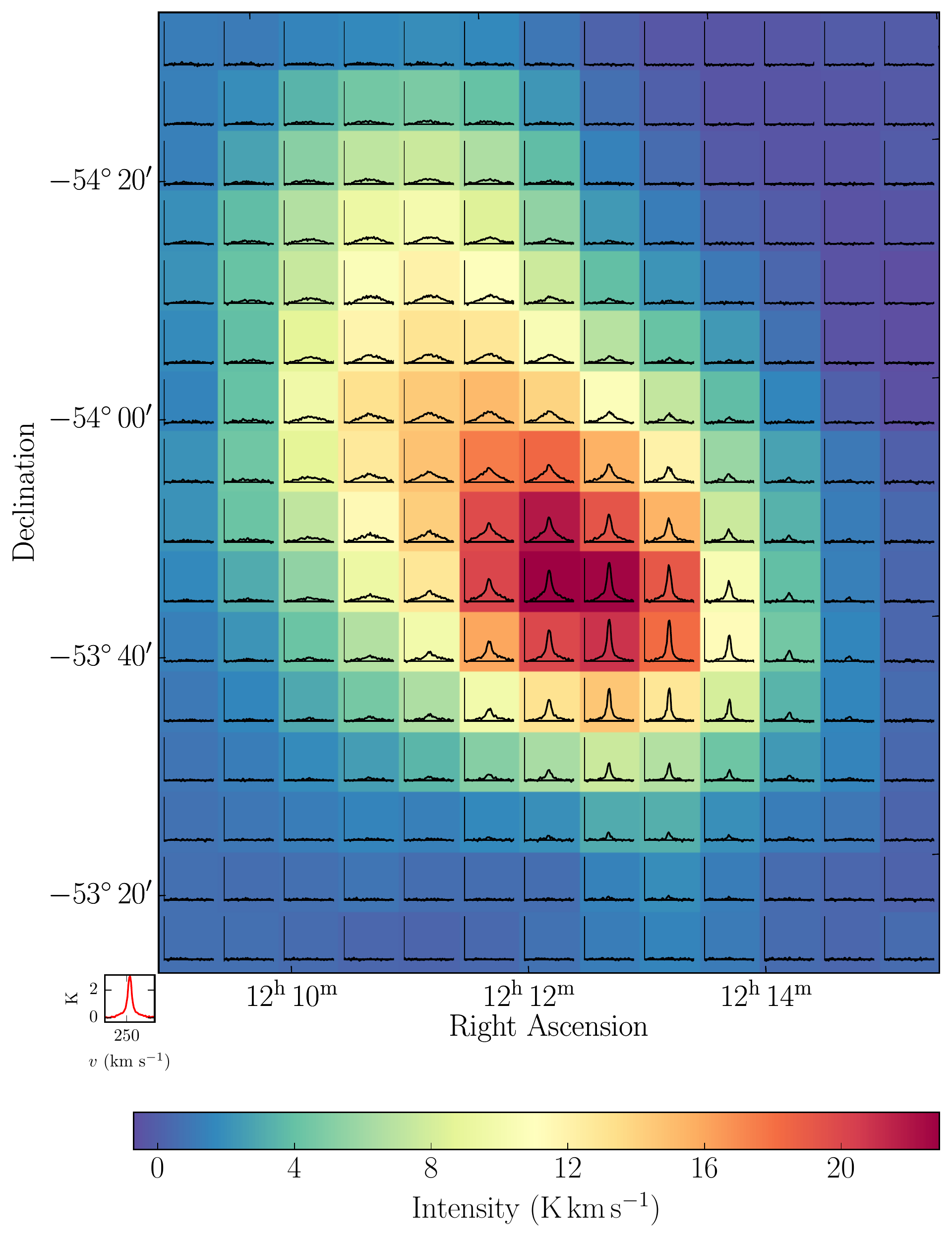}
  \includegraphics[width=0.45\textwidth, angle=0, trim=0 0 0 0]{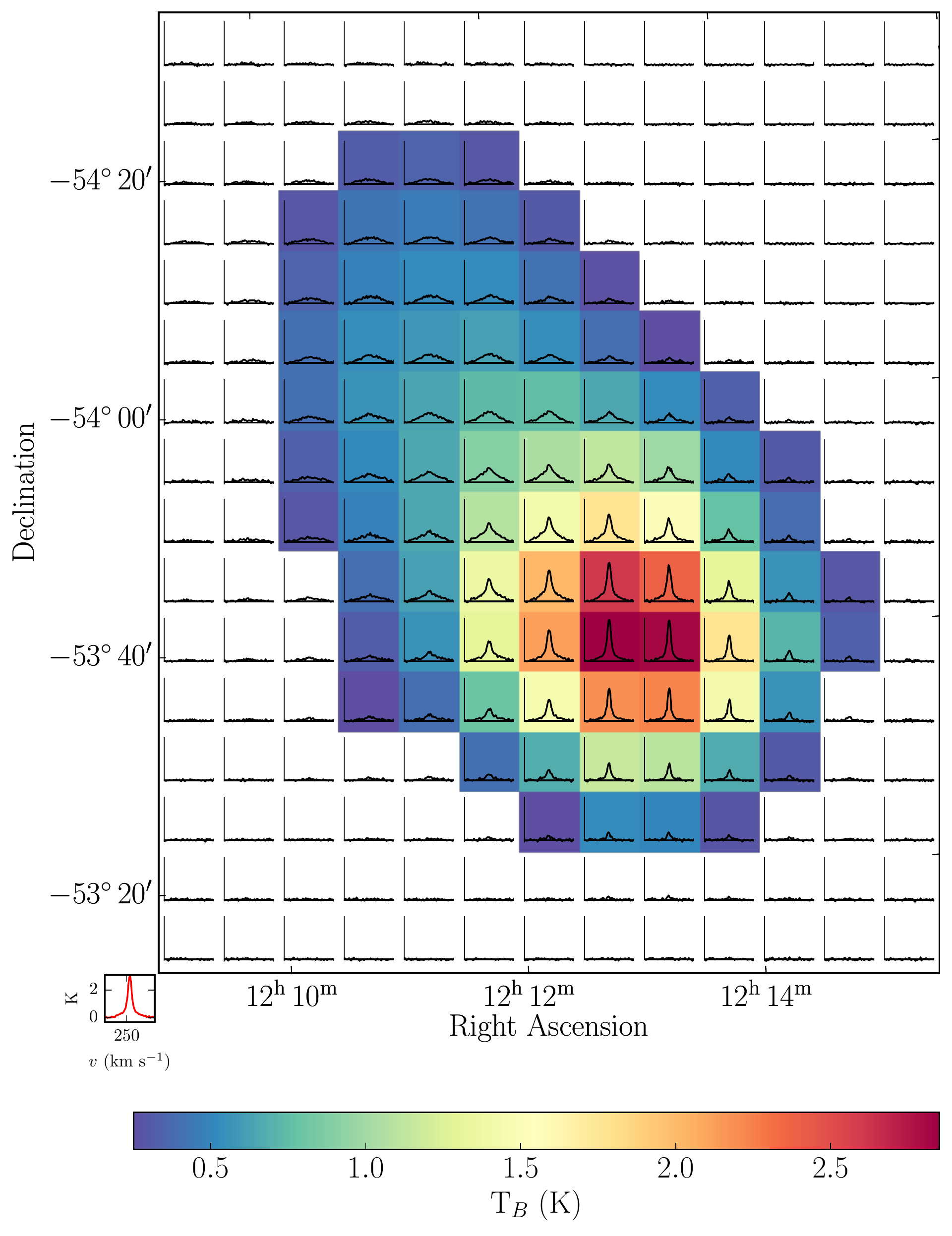}
        \includegraphics[width=0.45\textwidth, angle=0, trim=0 0 0 0]{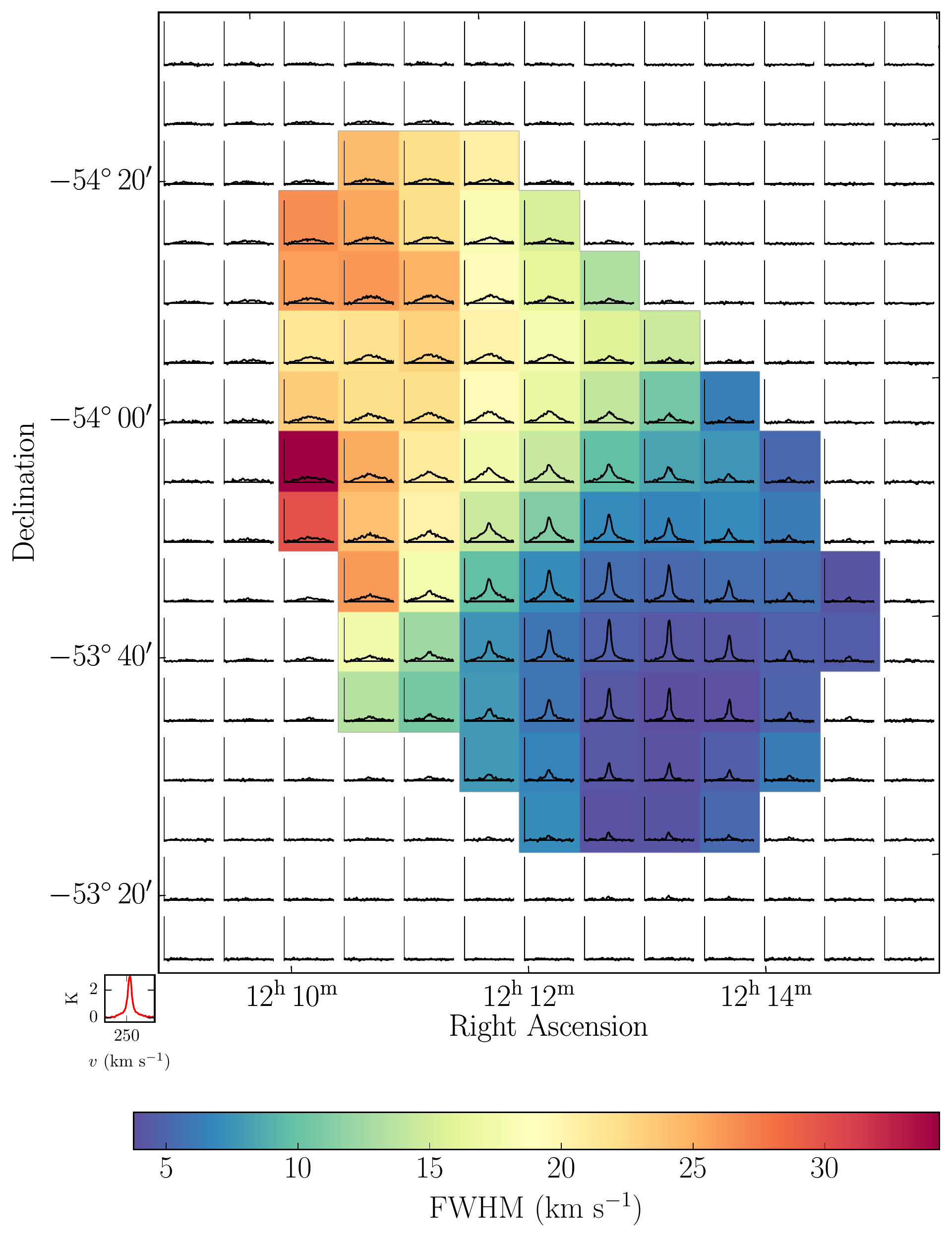}
    \includegraphics[width=0.45\textwidth, angle=0, trim=0 0 0 0]{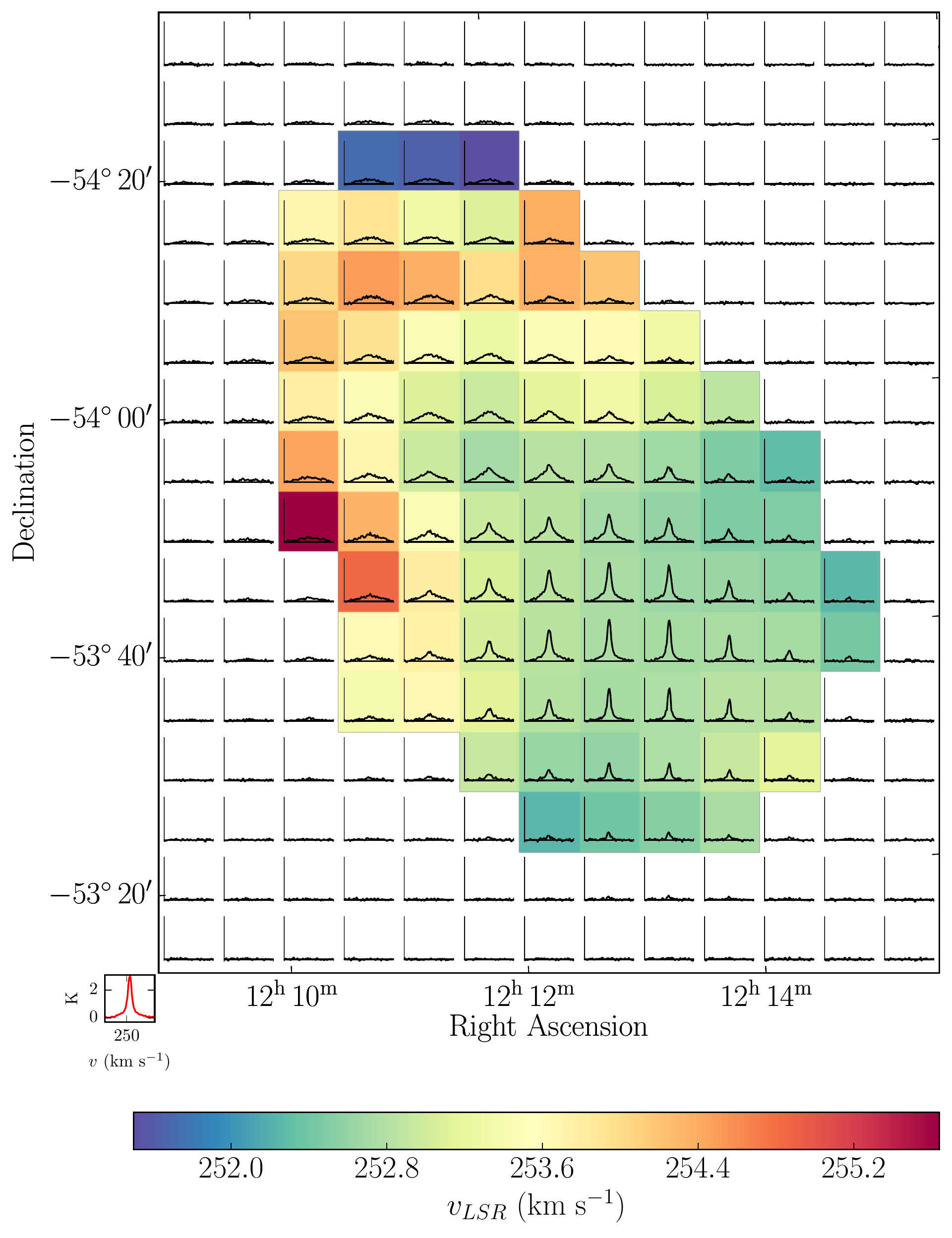}
  \caption{\small Different measured properties of catalogued cloud GHVC~G297.1+08.7+252 \citep{Moss:2013bu} displayed in colour for the four panels. The H{\scshape i} spectra of the cloud are overplotted on integrated intensity (upper left), peak brightness temperature (upper right), full-width at half-maximum (lower left) and velocity relative to the Local Standard of Rest (lower right). Each plot (with the exception of the integrated intensity) is masked at 320\,mK. Based on its physical structure, this cloud is argued to be ram-pressure stripped as it moves through the hot halo of the Milky Way \citep{BenBekhti:2006ef}, with bright, narrow components in the head and faint diffuse components in the tail.}
  \label{fig:g297}
\end{figure*}

In general, previous studies studying the distribution of cloud line widths have not revealed much in terms of useful discriminants between populations. For example, \citet{Saul:2012gt} reveal no evidence for a bimodal distribution in line widths within the GALFA-H{\sc i} Compact Cloud Catalog, \citet{Adams:2013dv} find no difference between line widths for Galactic CHVCs and UCHVCs from GALFA and \citet{Wolfe:2016fs} find similarly no difference between M31 HVCs and UCHVCs using the Green Bank Telescope. This is likely because of the difference in sensitivity, with most surveys not being able to probe the low column densities on a larger scale and compare them to the overall structure of clouds.

We present here a study of neutral atomic hydrogen gas at high velocity in the Milky Way, combining the very sensitive 21\,cm Green Bank survey for high-velocity H{\scshape i} \citep{Lockman:2002bu} with the search for high-velocity H{\scshape i} in the Galactic All Sky Survey \citep{Moss:2013bu}. These two surveys target significantly different aspects of the H{\scshape i} content of the Milky Way halo. \citet{Lockman:2002bu} sampled the halo with high sensitivity (median $\sigma$~=~3.4\,mK) at random positions (some towards extragalactic objects bright in optical and UV) with the 140\,ft (43\,m) NRAO telescope at Green Bank. \citet{Moss:2013bu} use automated source-finding to identify all southern-sky high-velocity clouds (HVCs) above a brightness of 4$\sigma$ ($\sim$230\,mK) in the Parkes 64\,m radio telescope Galactic All Sky Survey \citep[GASS,][]{McClureGriffiths:2009dn,Kalberla:2010jo}, which has an root mean square (RMS) noise of 57\,mK. For comparison, the \citet{Lockman:2002bu} data scales to an RMS of 13\,mK at the GASS spectral and spatial resolution. Together these different surveys reveal new insights into the structure of the neutral hydrogen in the halo of the Milky~Way. 

We outline the properties of each survey in Section \ref{surveys} and describe the sample used from the GASS catalogue of high-velocity clouds. We present our results and analysis in Section \ref{resultsdiff}, as well as a consideration of selection effects and survey limitations. Finally, we summarise our results and present avenues for future investigation in Section \ref{conclusiony}.

\section{Survey details and data}\label{surveys} 
\citet[][hereafter L02]{Lockman:2002bu} used the 140\,ft (43\,m) NRAO telescope at Green Bank to conduct a very sensitive search for faint high-velocity H{\scshape i} lines in 860 directions. The angular resolution of the data was 21$'$, with a velocity resolution of 5\,km~s$^{-1}$ over the $v_{LSR}$ range $-$1000\,km~s$^{-1}$ to $+$800\,km~s$^{-1}$. The median RMS sensitivity of the pointed observations was 3.4\,mK, which gives a median 4$\sigma$ detection limit of $N_{HI} \sim$ 8 $\times$ 10$^{17}$\,cm$^{-2}$ (for an assumed typical line-width of 30\,km~s$^{-1}$). This sensitivity is much higher than large-scale surveys of Galactic H{\scshape i} and offers unique insight into the faint high-velocity Galactic halo that has generally only been sampled effectively by UV absorption lines \citep[e.g.][]{Savage:1996dz,Wakker:2003fx,Lehner:2012fc}. Part of the motivation of the study was to investigate the incongruity between the amount of high-velocity neutral hydrogen revealed by current H{\scshape i} surveys and the amount that should be present based on the results from UV absorption studies. As such, one third of the observations were made specifically towards extragalactic objects visible at optical and infrared wavelengths in order to facilitate future follow-up. L02 detected high-velocity H{\scshape i} in 37\% of their observed sightlines, with the detected lines generally found to be associated with known high-velocity complexes (such as the Magellanic system and Complex C) and typically extended over several degrees rather than being compact in structure. They concluded based on these results that most HVCs do not have sharp edges, and that the high-velocity H{\scshape i} detected in their survey was likely to be associated with the faint edges of HVC complexes not detectable by previous H{\scshape i} surveys.

The Galactic All Sky Survey \citep{McClureGriffiths:2009dn} was made using the 21\,cm multibeam receiver on the Parkes 64\,m radio telescope to survey the entire southern sky in neutral hydrogen, complete over the Galactic velocity range of $v_{LSR}$ = $\pm$468\,km~s$^{-1}$. It features an effective angular resolution of $\sim$16$'$, a spectral resolution of 1\,km~s$^{-1}$ and an average RMS sensitivity of 57\,mK. The combination of high angular resolution, high sensitivity and high spectral resolution found in GASS makes it an excellent southern-sky survey of Galactic neutral hydrogen, particularly of relatively cold gas with narrow line-widths throughout the Milky Way as well as relatively faint gaseous structures in the Galactic halo. 

\begin{figure}[t]
  \centering
  \includegraphics[width=0.48\textwidth, angle=0, trim=0 0 0 0]{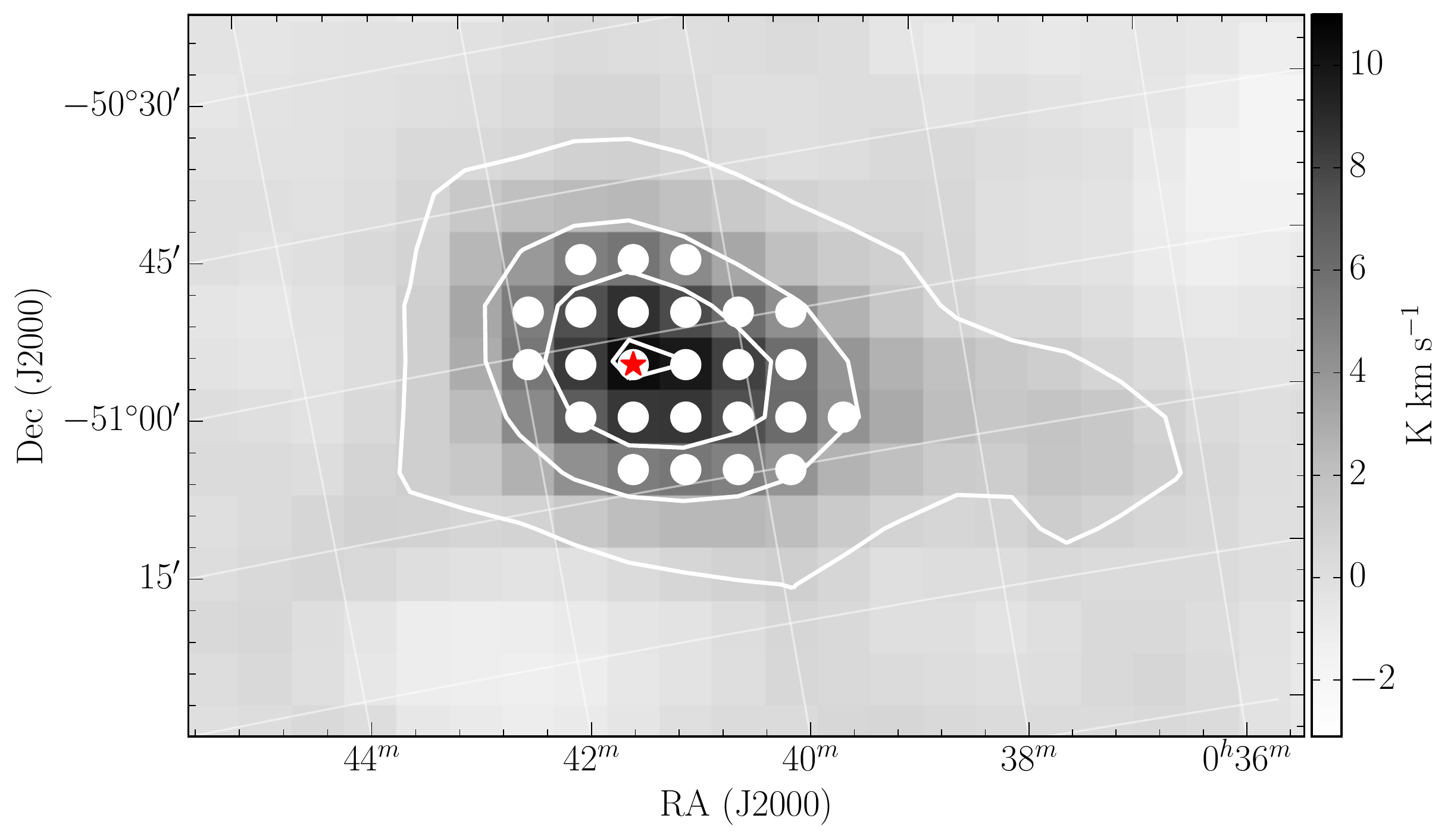}
  \caption{\small Example of the distribution of spectral components across a single GASS HVC, GHVC~G306.9$-$66.0+166. Each white point indicates a position at which the spectrum through the source contains a peak brightness temperature $T_B$ $>$ 5$\sigma$. The red star indicates the catalogued position (cloud core) as in \citet{Moss:2013bu}. The background is the integrated intensity image of all velocity slices of the cloud which contain emission brighter than 2$\sigma$ (0.11\,K), with white contours (4 equally-spaced levels from 25\% to the maximum of the image) overlaid of the same cloud to aid in highlighting the source.}
  \label{fig:g307}
\end{figure}

The GASS catalogue of HVCs \citep{Moss:2013bu} was constructed using automated source-finding on the Galactic All Sky Survey second release \citep[GASS~II,][]{Kalberla:2010jo}. We refer to the GASS~II data as GASS throughout this paper. We also note that both L02 and GASS are based on an absolute TB scale consistent with the standard sources S6 and S8, and were also corrected for stray radiation. To identify clouds, \citet{Moss:2013bu} adopted an implementation of the floodfill source-finding algorithm developed by \citet{Murphy:2007uq} extended to a three-dimensional cube. In this catalogue, \citet{Moss:2013bu} identified 1111 positive velocity HVCs and 582 negative velocity HVCs in the southern sky with a peak brightness temperature $T_B >$ 4$\sigma$ ($\sim$230\,mK). The catalogue also included 295 anomalous velocity clouds (AVCs) which did not meet the traditional velocity criteria (peak deviation velocity $>$ 50\,km~s$^{-1}$ and local standard of rest (LSR) velocity $>$ 90\,km~s$^{-1}$) to be classified as an HVC. The sensitivity of GASS is less than that of the H{\scshape i} targeted with the very sensitive Green Bank survey of \citet{Lockman:2002bu}, and as such the GASS catalogue of HVCs selects for relatively bright high-velocity H{\scshape i} clouds. However, the sensitivity to large-scale structure of GASS compared to previous southern-sky surveys of H{\scshape i} is particularly important in the context of HVCs as bright clumps connected by fainter diffuse gas, meaning that it was possible to identify complexes of clouds connected by faint gas as well as more isolated clumps of gas down to a threshold limit of 2$\sigma$ ($\sim$110\,mK).

\subsection{Sampling spectra across HVCs in GASS}\label{samples}
With GASS, it is possible to investigate how spectral structure (and hence physical properties) change across individual clouds. An example of spectral variation across HVCs for GHVC~G297.1+08.7+252 is shown in Figure \ref{fig:g297}, showing integrated intensity, peak brightness temperature ($T_B$), full-width at half-maximum (FWHM) and velocity relative to the Galactic Standard of Rest ($v_{GSR}$). This particular cloud has been argued to be ram-pressure stripped as it moves through the hot halo of the Milky Way producing a head-tail structure\citep{BenBekhti:2006ef}, with a concentration of bright, narrow, cold components in the head and broad, faint, warm components in the tail. Head-tail HVCs like GHVC~G297.1+08.7+252 are not unusual, but are found all over the Galactic halo with distinct column density and velocity gradients \citep{Putman:2011gra}. These clouds show clear evidence of interacting with the surrounding halo medium, and allow us to probe both their physical conditions and environment through their disrupted spectral properties.

\begin{figure}[t]
  \includegraphics[width=0.48\textwidth, angle=0, trim=0 0 0 0]{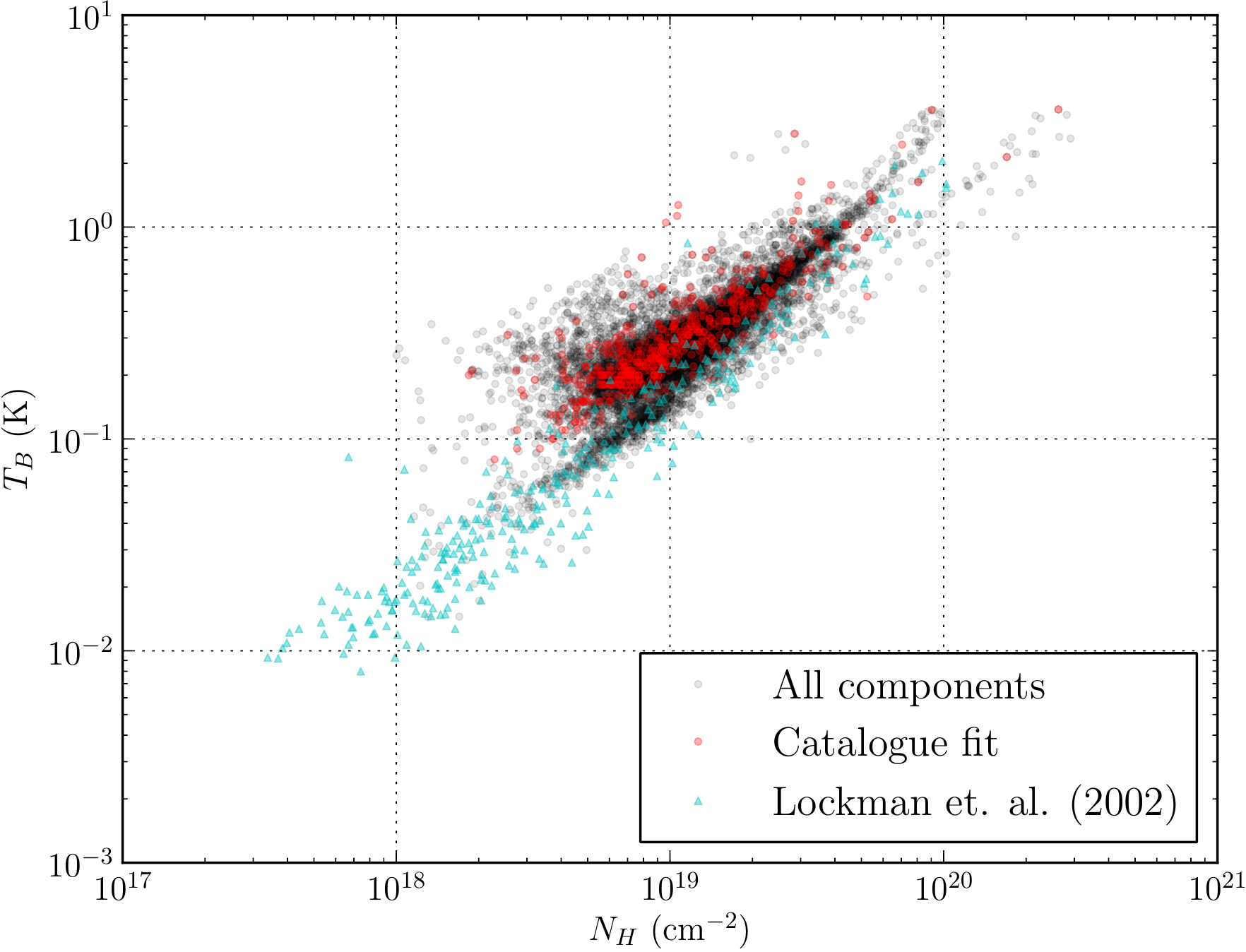}
  \includegraphics[width=0.48\textwidth, angle=0, trim=0 0 0 0]{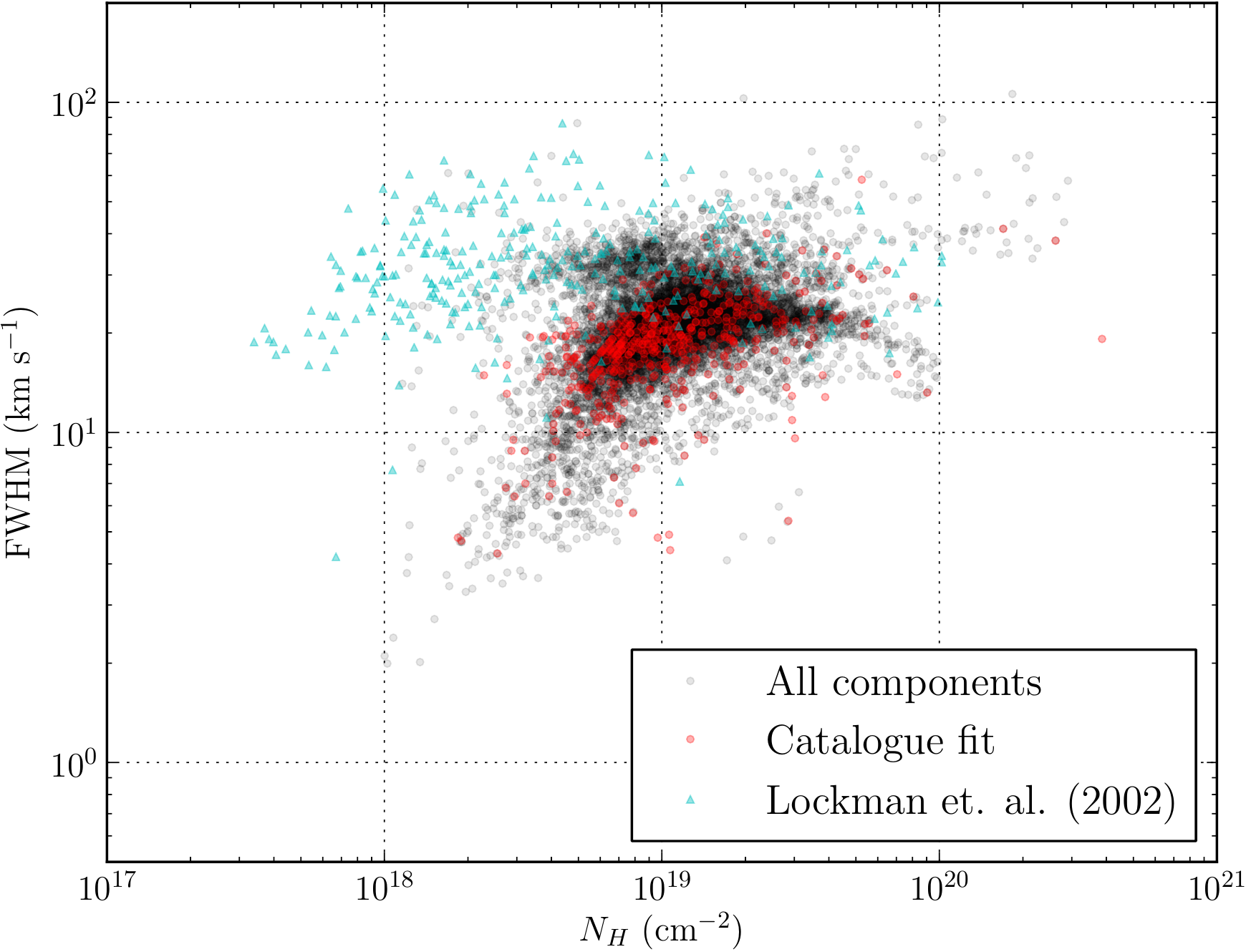}
  \caption{\small Scatter plots of both the GASS HVC population and the L02 population, for $T_B$ vs. $N_{HI}$ (top) and FWHM vs. $N_{HI}$ (bottom). The properties at the cloud cores (catalogue fit) of the GASS HVCs used in our sample are shown as red circles, while the spectral components across the clouds are shown as black circles. Overlaid are the properties of the L02 population as cyan triangles. In both cases, we see that the GASS HVC spectral components are clearly divisible into two populations: a minority of components which align with the L02 distribution, and the majority of components which are brighter and narrower in line-width.}
  \label{fig:allcomps}
\end{figure}

\begin{figure}[t]
  \includegraphics[width=0.48\textwidth, angle=0, trim=0 0 0 0]{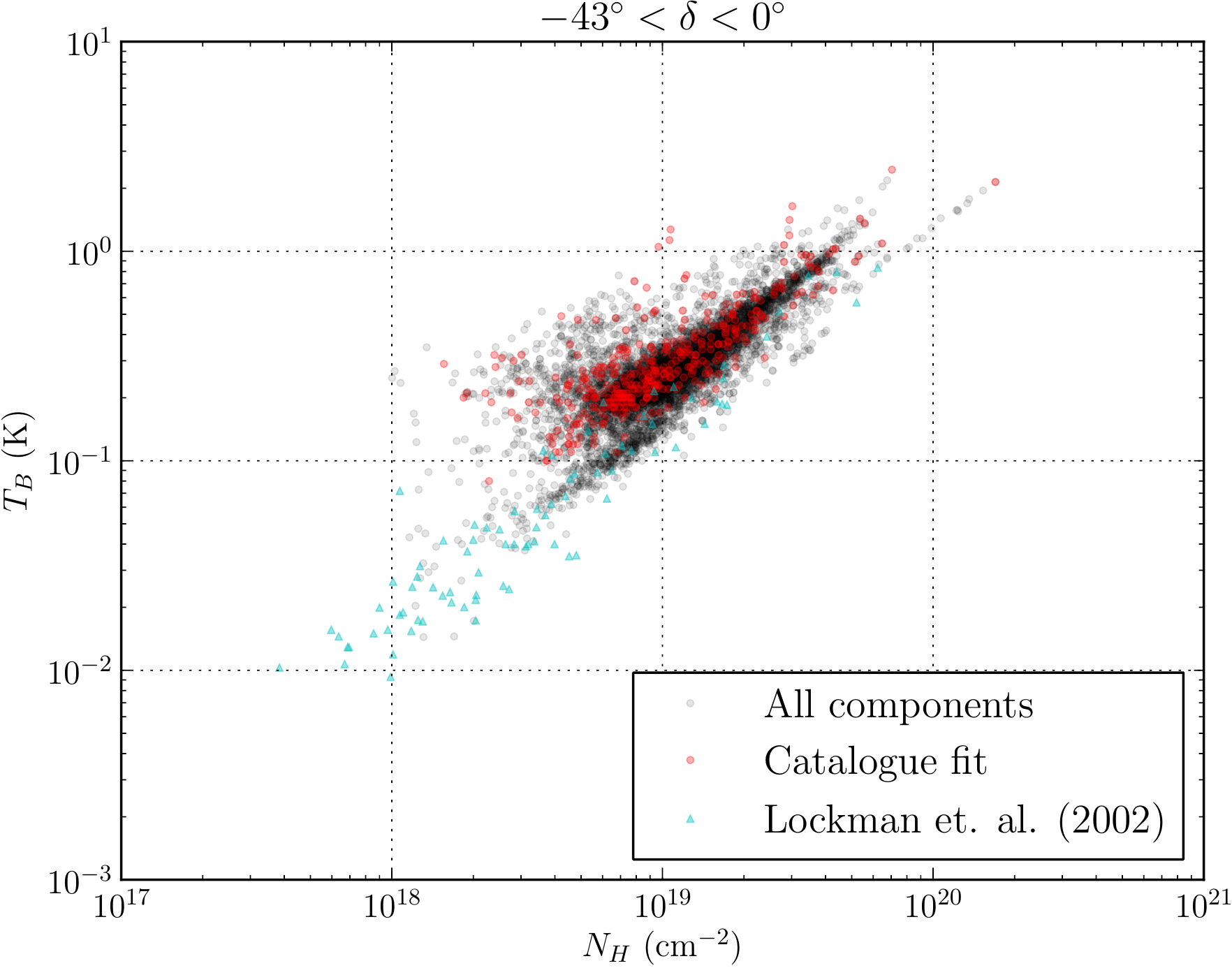}
  \includegraphics[width=0.48\textwidth, angle=0, trim=0 0 0 0]{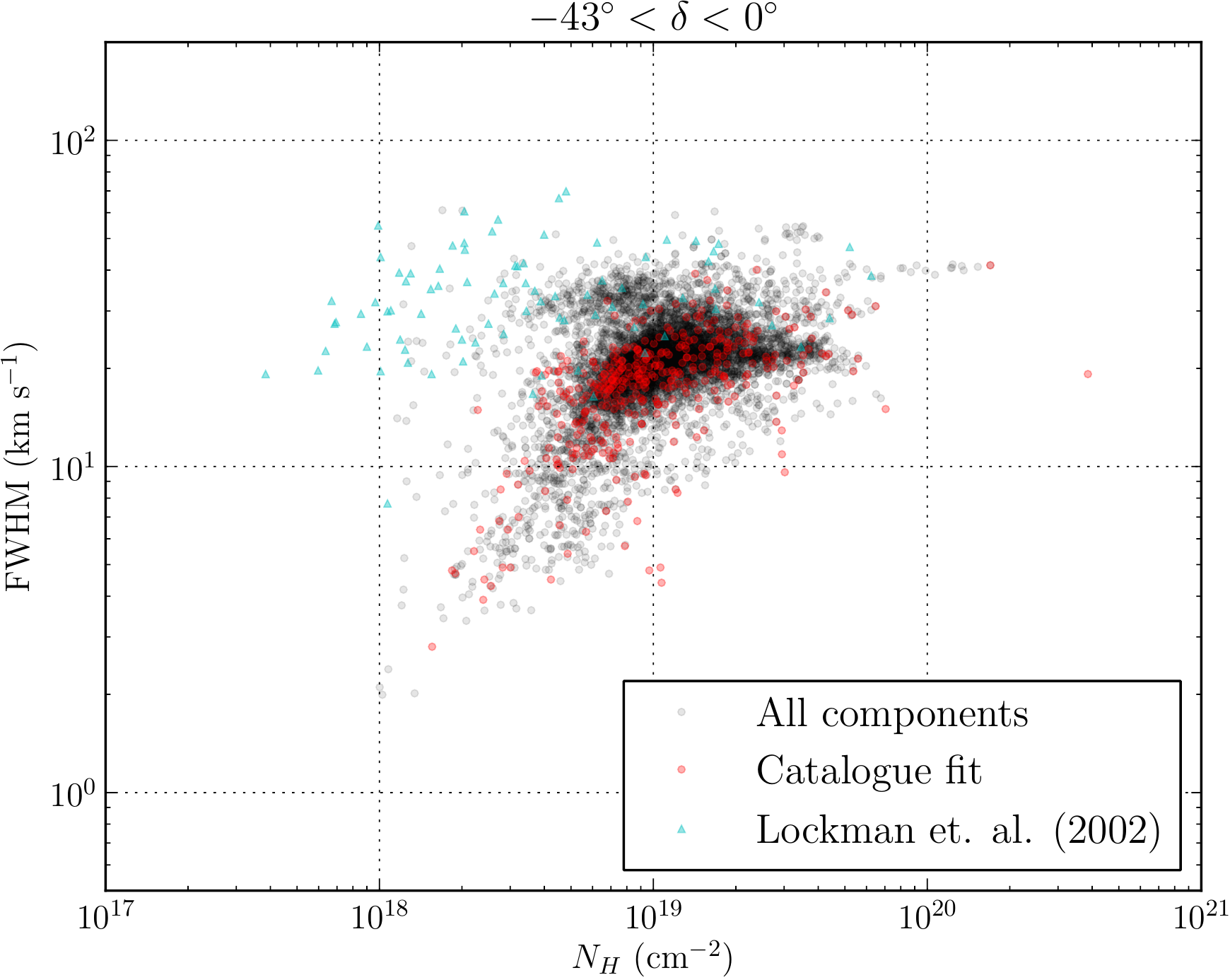}
  \caption{\small Scatter plots of both the GASS HVC population and the \citet{Lockman:2002bu} population, in $T_B$ vs. $N_{HI}$ (top) and FWHM vs. $N_{HI}$ (bottom), in only the declination overlap region between the two surveys of $-$43$^\circ < \delta <$ 0$^\circ$. The same trends as seen in the entire sample are seen in this directly comparable subset, confirming our result and negating potential influence of hemispherical selection effects.}
  \label{fig:overlap}
\end{figure}

Here we use the spectral sample as described in \citet{Moss:2013bu}: we obtain all spectral components with a peak brightness temperature above 5$\sigma$ (285\,mK in GASS) in all clouds with a total size $<$ 5 deg$^2$. Considering only the high-velocity gas to be that with $|v_{LSR}|$ $\ge$ 90\,km~s$^{-1}$, we obtained a total of 12210 spectra across a total of 549 GHVCs. Each spectrum corresponds to a pixel, where the pixel size is 4.8$'$ for a total of 3.33 pixels across the $\sim$16$'$ beam. For the purposes of this analysis, we consider only a single-component fit to each spectrum as a means of differentiation between broad and narrow lines, with the goal that a spectrum dominated by either a narrow component or a broad component will be characterised as such by the single-component fit. An example in the case of a individual cloud, GHVC~G306.9$-$66.0+166, is shown in Figure \ref{fig:g307}, where the white points indicate positions at which spectra were included in the sample.

We use the following criteria on this sample of spectra:
\begin{enumerate}
\item We obtain the column density for each cloud using 
\begin{equation}\label{columndenseq}
N_{HI} = 1.823 \times 10^{18} \sqrt{\frac{\pi}{4 \ln 2}}~T_{B,fit}~\Delta v~\textrm{cm}^{-2},
\end{equation}
where $T_{B,fit}$ is the brightness temperature in K and $\Delta v$ is the line-width in km~s$^{-1}$, and then ignore any spectral components with a column density $N_{HI} <$ 10$^{18}$ cm$^{-2}$ as this is below the sensitivity limit of the GASS data. This sensitivity limit was calculated assuming a 4$\sigma$ detection (0.228\,K) and a line-width of 3 channels (2.4\,km~s$^{-1}$). While all spectra with $N_{HI} <$ 10$^{18}$ cm$^{-2}$ were flagged, we kept their host clouds in the sample. However, we did remove 12 clouds that featured $>$ 30\% of spectral components with a column density $N_{HI} <$ 10$^{18}$ cm$^{-2}$, as these were found to be located in noisy regions or dominated by noise spikes.

\item Any clouds contributing $>$ 300 components across the cube (corresponding to 1.9\,deg$^2$) are removed from the sample. This filter applied to:\\
GHVC~G222.2$-$87.9$-$263, \\
GHVC~G280.6$-$05.6+186, \\
GHVC~G075.8$-$73.6$-$121.\\ 
Clouds like these contributing a few hundred components are located close to the Magellanic emission or to Galactic emission, and are not representative of the general cloud population.

\end{enumerate}

Upon applying these filters to our data, we obtain a total of 10455 usable spectra across 534 clouds.

\section{Dense and diffuse neutral hydrogen in the halo}\label{resultsdiff}
Figure \ref{fig:allcomps} shows the results of combining the L02 sample with our sample of spectra across GASS HVCs as described in Section \ref{samples}, for both brightness temperature $T_B$ and FWHM versus column density $N_{HI}$. The properties at the catalogued position (the core of the HVC, see \citet{Moss:2013bu} for method of determining this position) of each GASS HVC are plotted as a red circle, while the population of all spectral components is plotted as black circles. For comparison, we plot the L02 distribution of spectral components as cyan triangles. In this case, there are 10455 GASS spectral components plotted, 534 GASS cloud cores plotted and 305 L02 spectral components plotted. 

\begin{figure}[t]
  \includegraphics[width=0.48\textwidth, angle=0, trim=0 0 0 0]{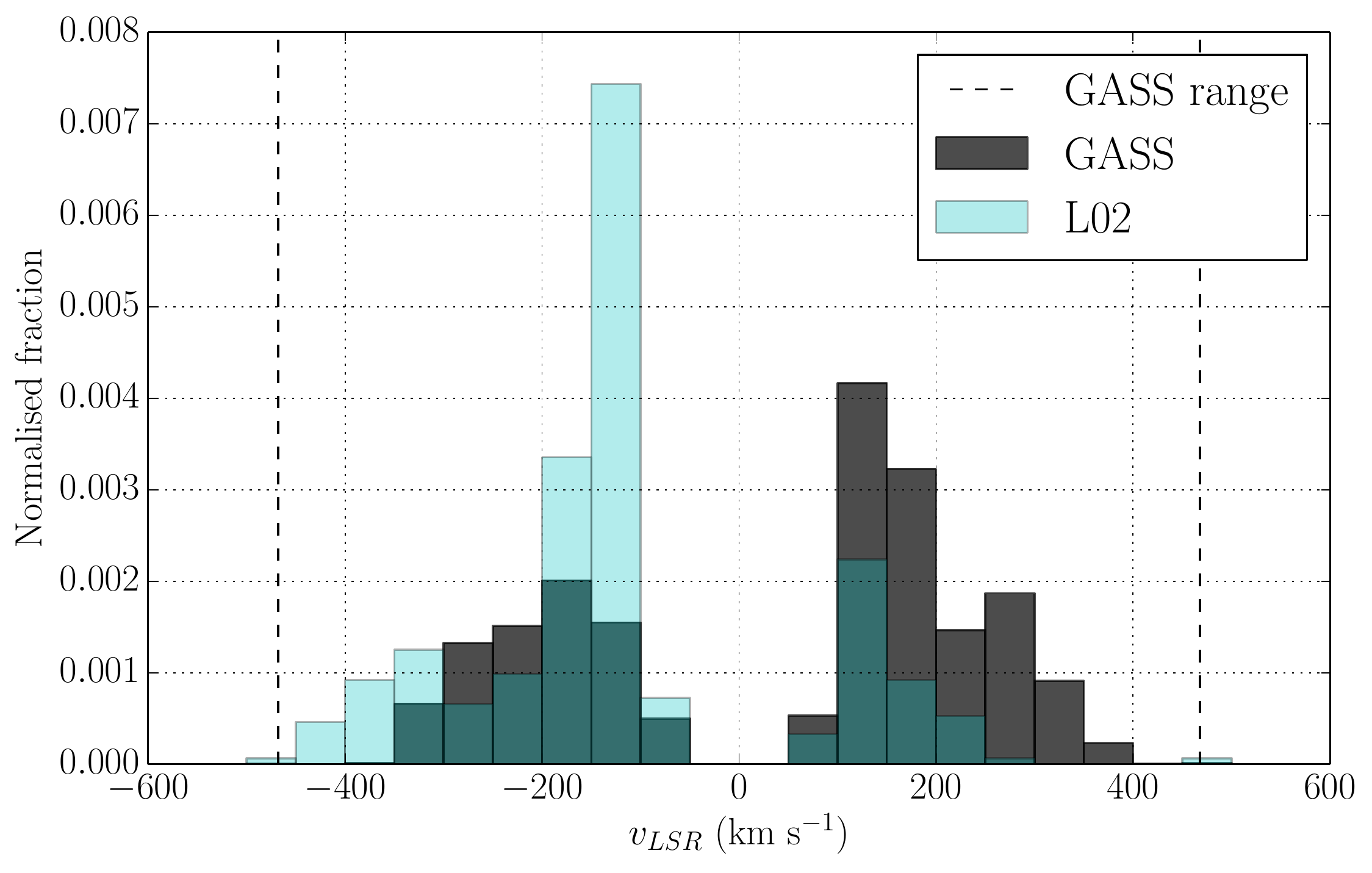}
  \caption{\small Comparison of the velocities of detected clouds in GASS and L02. Shown is the distribution of all 10455 GASS spectral components in black and all 305 L02 components in cyan. The published velocity coverage of GASS ($v_{LSR}$ = $\pm$468\,km~s$^{-1}$) is shown as two dashed lines. This highlights that the extended velocity coverage of L02 does not result in any significant difference in the observed HVC population, as the majority of HVCs detected in both surveys fall within the GASS velocity limits. The dominance of negative velocities in L02 and positive velocities in GASS is due to rotation of the Milky Way \citep{Wakker:1991ww}. The histogram is normalised to enable comparability due to the large number of GASS components.}
  \label{fig:velcov}
\end{figure}

\begin{figure*}[t]
\centering
   \includegraphics[width=0.45\textwidth, angle=0, trim=0 0 0 0]{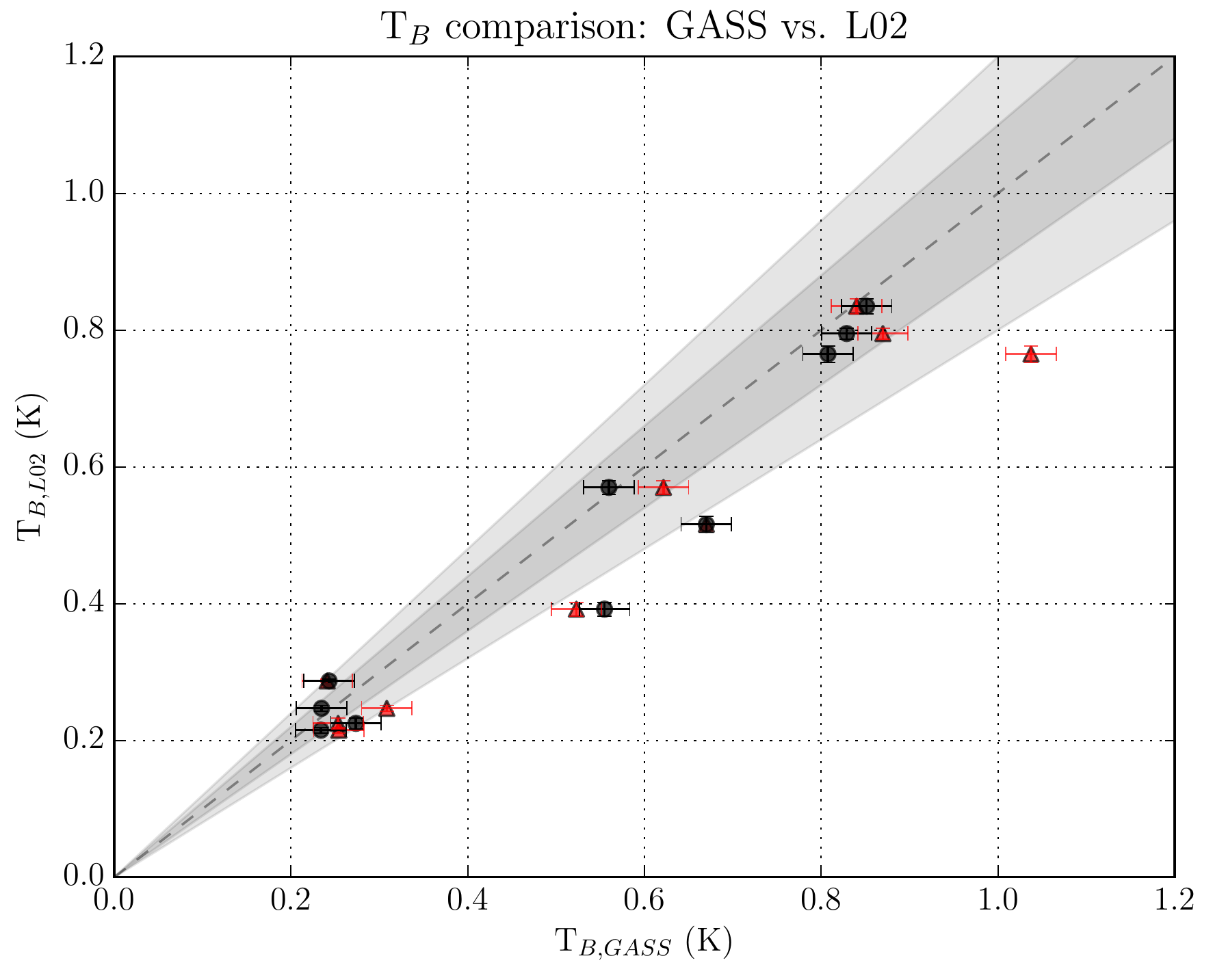}
       \includegraphics[width=0.47\textwidth, angle=0, trim=0 0 0 0]{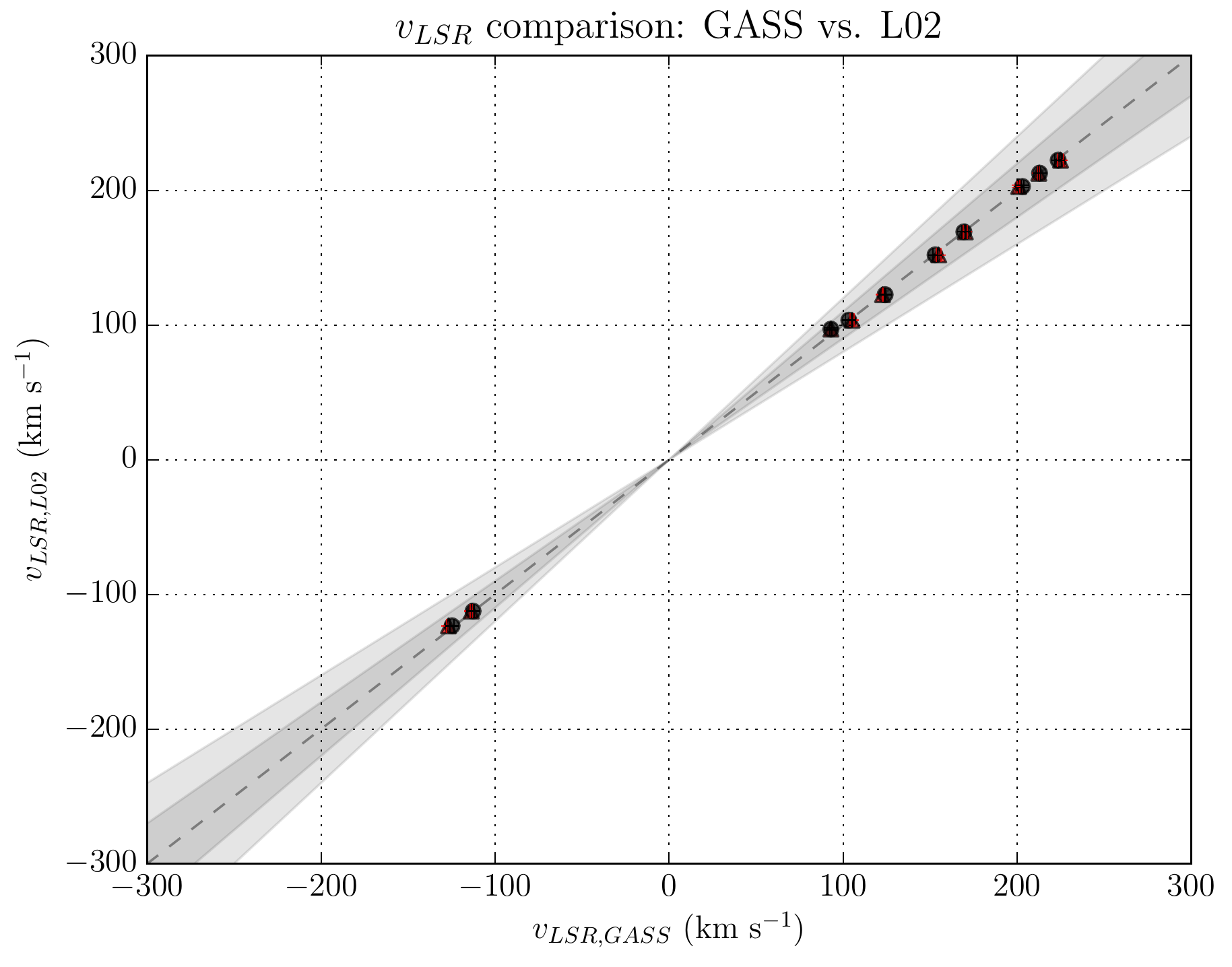}
    \includegraphics[width=0.46\textwidth, angle=0, trim=0 0 0 0]{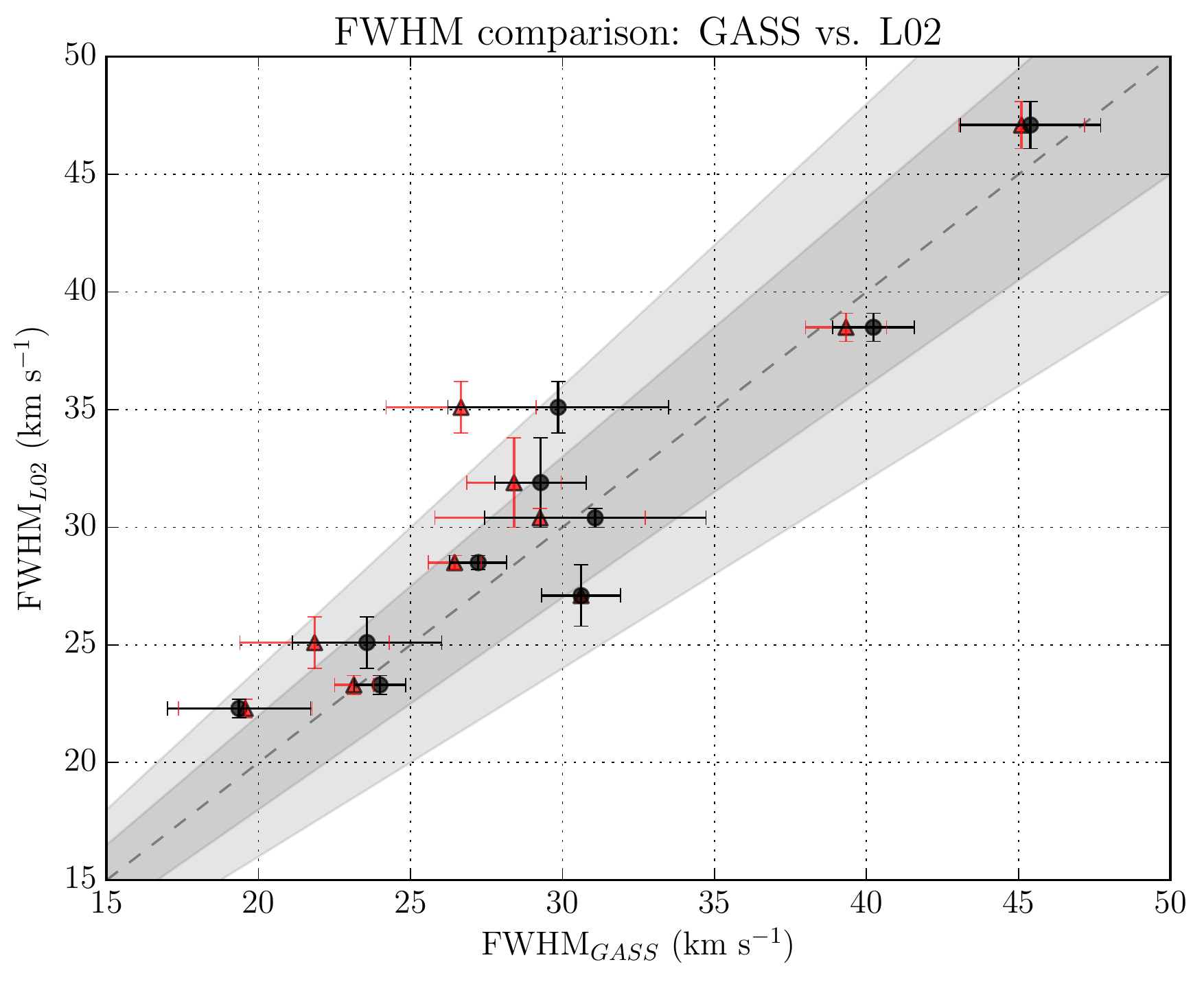}
  \includegraphics[width=0.45\textwidth, angle=0, trim=0 0 0 0]{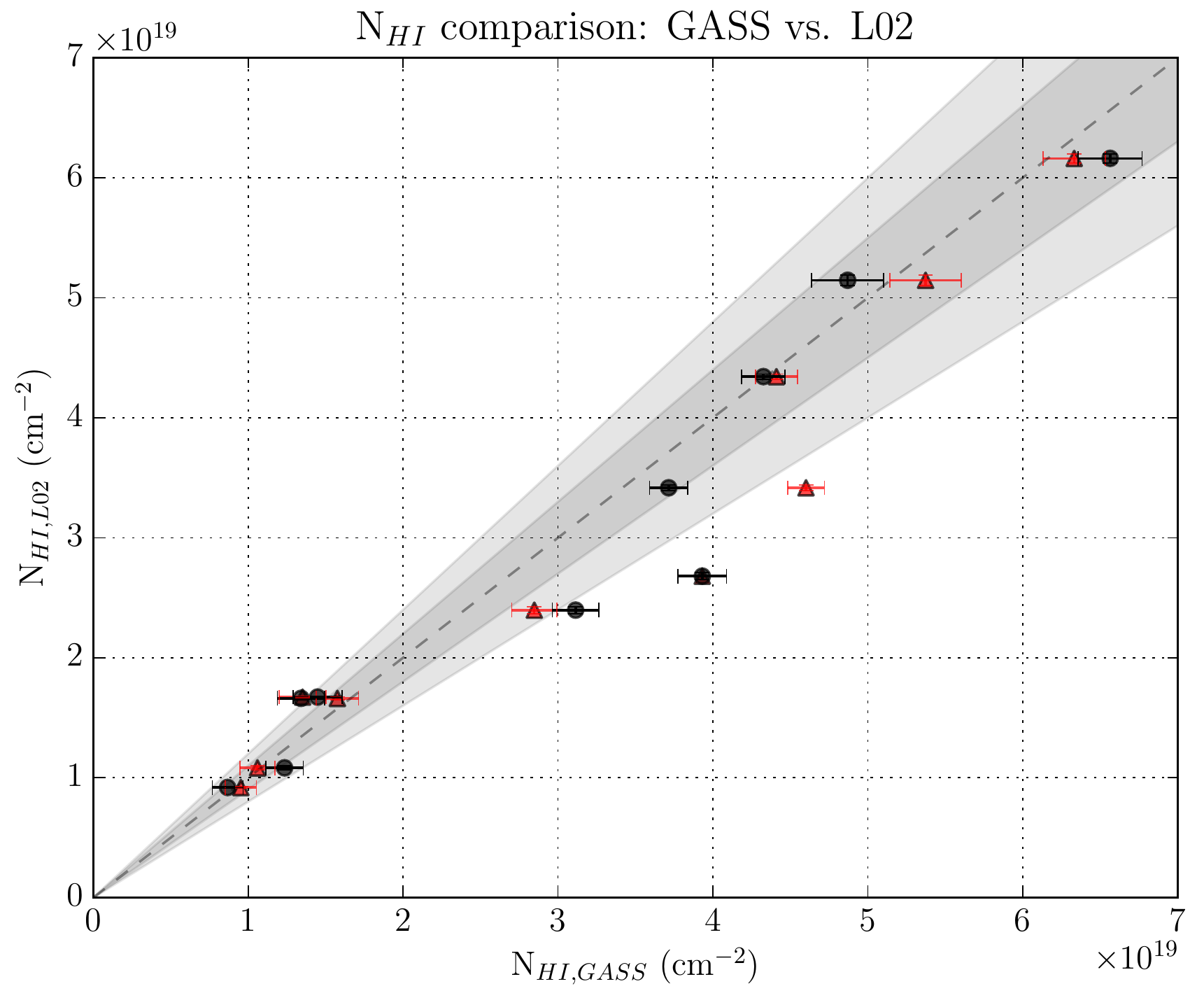}
  \caption{\small Comparison of measured properties for L02 sightlines visible in GASS. A total of 10 L02 sightlines are both detectable and comparable with GASS, plotted on each graph above. The compared properties are peak brightness temperature $T_B$ (top left), velocity $v_{LSR}$ (top right), line-width FWHM (bottom left) and column density $N_{HI}$ (bottom right). The grey shaded areas are $\pm$10\% and 20\% of the dashed line, which represents the 1:1 ratio. The black circles represent measurements performed on GASS data smoothed to 21$'$ spatial resolution and 5 km s$^{-1}$ velocity resolution to match L02, while the red triangles show the measurements on the unsmoothed GASS data. Overall, the velocity shows the most agreement, while there is slight deviation in the other properties.}
  \label{fig:lockgass}
\end{figure*}

What is evident in both plots is a clear distinction between two types of spectra that separate at a given column density. At each step in column density, there is a bright narrow population typical of the majority of GASS HVC components while the L02 data traces a population of H{\scshape i} that is fainter and of broader line-width. We henceforth refer to the two evident populations as the dense population and the diffuse population, respectively. 

\subsection{Selection effects and sample~limitations}\label{selection}
To examine possible systematic effects between the two surveys, which have significantly different sky coverage and observational parameters, we reproduce the earlier $T_B$($N_{HI}$) and FWHM($N_{HI}$) correlations in Figure \ref{fig:overlap}, using data only from the part of the sky covered by both ($-$43$^\circ < \delta <$ 0$^\circ$) which we refer to as the overlap region. In this case, there are 6851 GASS spectral components plotted, 394 GASS cloud cores and 83 L02 spectral components plotted. These spectra in the overlap region represent 66\%, 74\% and 27\% of the full sample, respectively. This figure shows the same overall trend as Figure \ref{fig:allcomps}: most GASS HVC lines cluster near the GASS cloud cores, with a branch of the spectral components that aligns with the brightest parts of the L02 distribution. This confirms our result for the full sample, and suggests that the distinction applies to both the northern and southern hemisphere sky. 

We also considered the effect of velocity coverage on the sample obtained. As mentioned previously, GASS is complete within LSR velocities of $\pm$468\,km~s$^{-1}$ while L02 covered the $v_{LSR}$ range $-$1000\,km~s$^{-1}$ to $+$800\,km~s$^{-1}$. The results are shown in Figure \ref{fig:velcov}, with the GASS spectral components in black and the L02 components in cyan. In both surveys, the majority of observed HVCs are within the GASS coverage of $v_{LSR}$ = $\pm$468\,km~s$^{-1}$. This indicates that despite the more extended velocity coverage of L02, there is no kinematic difference in the detected HVC populations. 

\begin{figure*}
\centering
\includegraphics[width=0.45\textwidth, angle=0, trim=0 0 0 0]{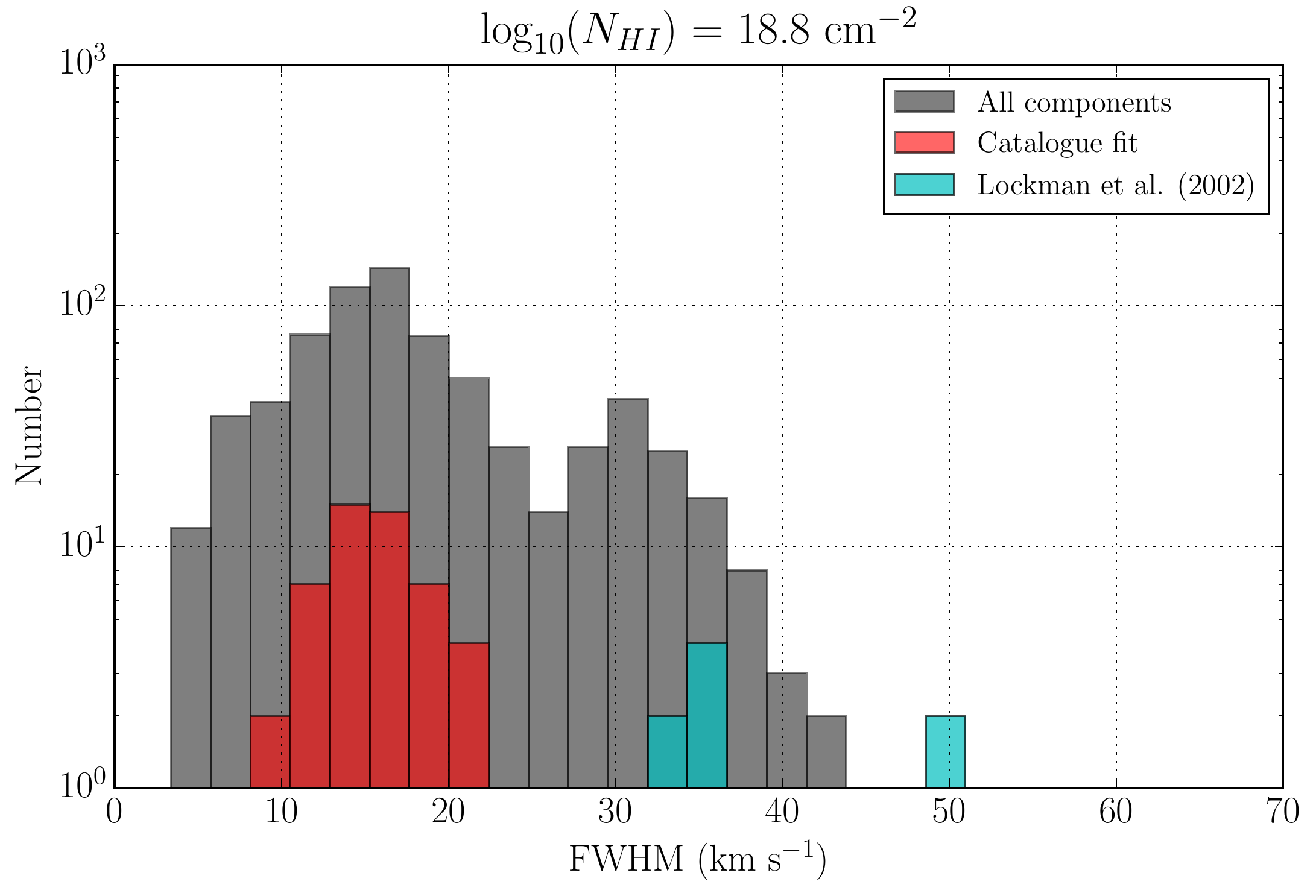}
\includegraphics[width=0.45\textwidth, angle=0, trim=0 0 0 0]{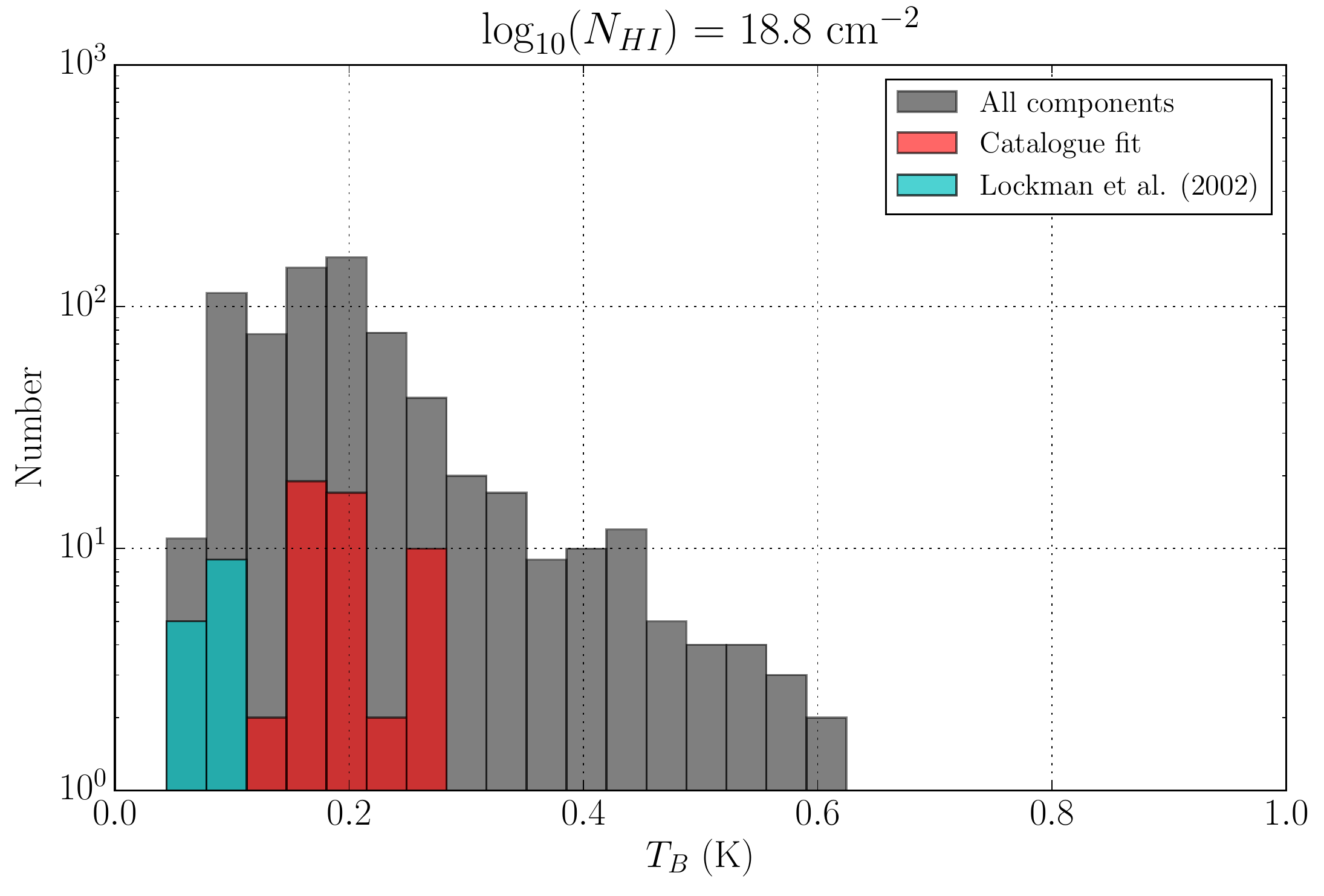}
\includegraphics[width=0.45\textwidth, angle=0, trim=0 0 0 0]{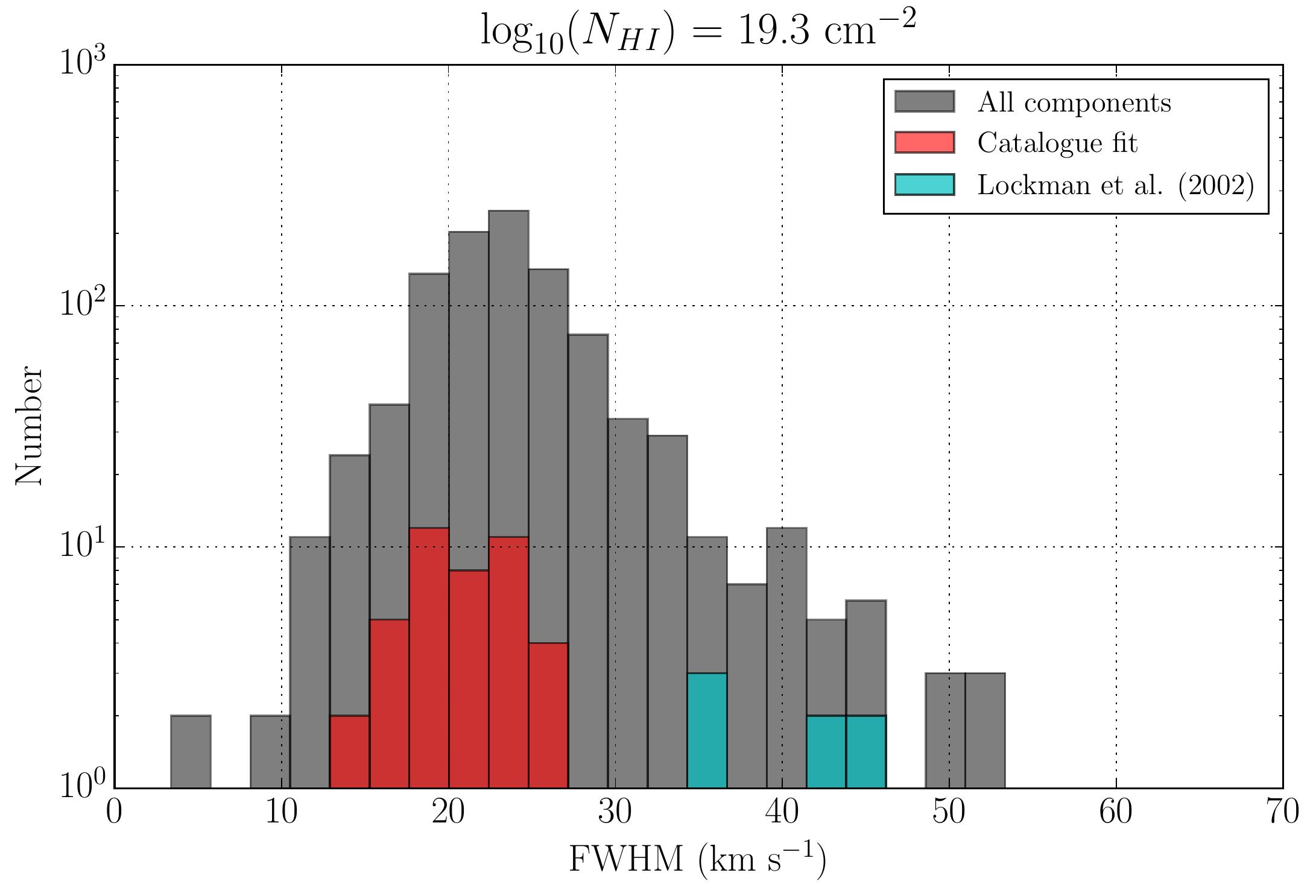}
\includegraphics[width=0.45\textwidth, angle=0, trim=0 0 0 0]{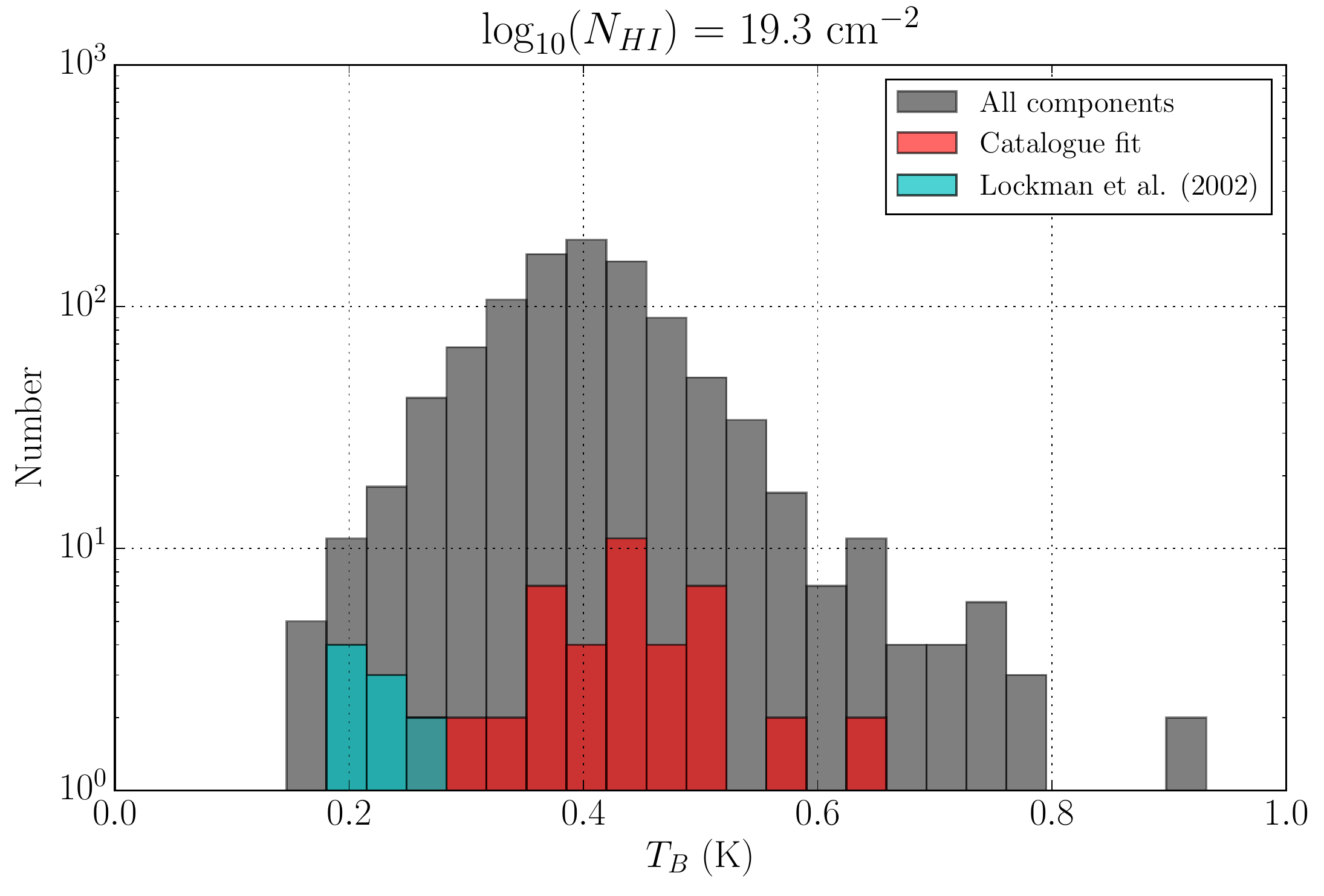}
\caption{\small Example histograms showing the distribution of the different populations for both FWHM and $T_B$, generated in steps of 0.1 in $\log_{10}(N_{HI})$. The GASS spectral components (grey) form a background distribution with respect to the GASS HVC cloud cores (red) and L02 components (cyan). Our goal with the machine learning approach is to divide the GASS components as best as possible into a population which agrees with the GASS cores (dense) and a population which agrees with L02 (diffuse), and thus gain further insight into the two populations.}
  \label{fig:gasshist}
\end{figure*}

\begin{figure*}[t]
\centering
\includegraphics[width=0.95\textwidth, angle=0, trim=0 0 0 0]{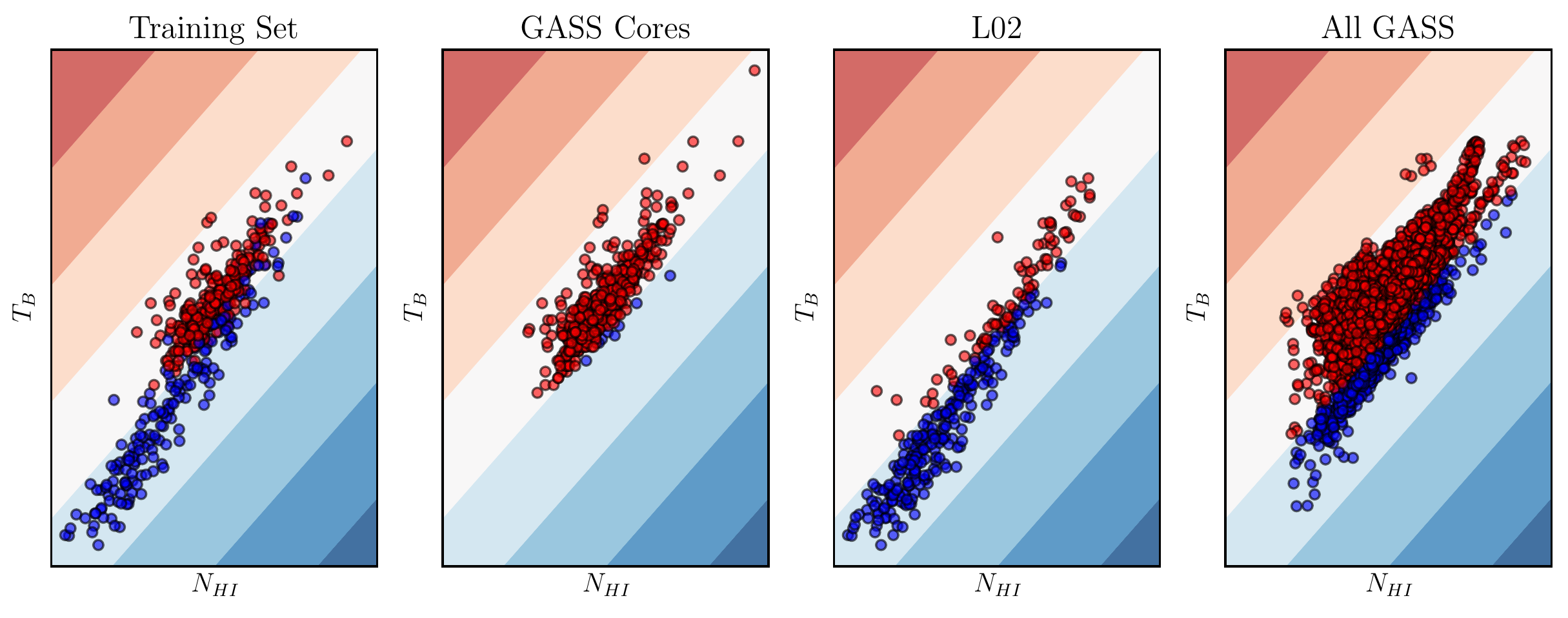}
\includegraphics[width=0.95\textwidth, angle=0, trim=0 0 0 0]{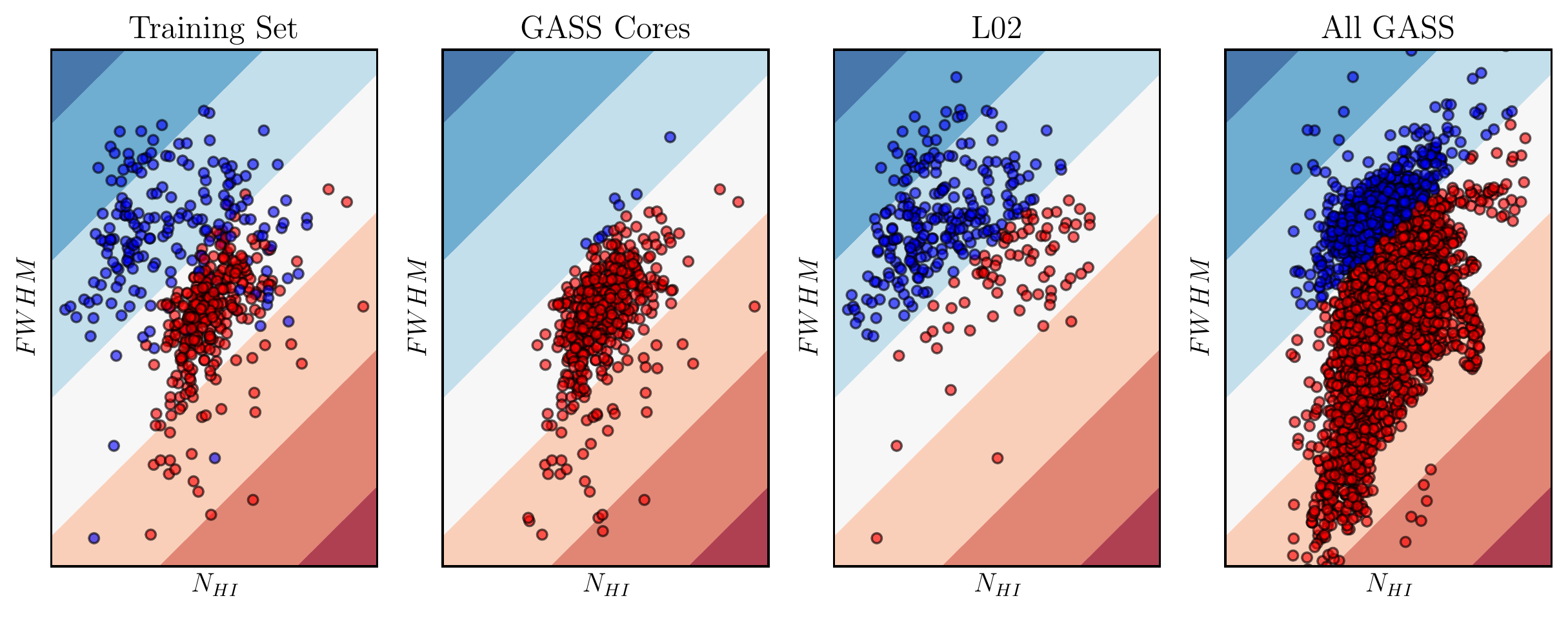}
\caption{\small Linear discriminant analysis (LDA) results in 2D for $f$($T_B$, $N_{HI}$) and $f$(FWHM, $N_{HI}$). The first panel shows the training subset (60\%) taken from the combined sample of GASS cores and L02, while the middle panels show the result of applying the classifier to these two complete sets after training. The far right panel shows the classification results for the entire GASS sample. The quantitative results are given in Table \ref{tb:mlcompare}.}
  \label{fig:ml2d}
\end{figure*}

\begin{figure*}[t]
\centering
\includegraphics[height=0.4\textwidth, angle=0, trim=0 0 0 0]{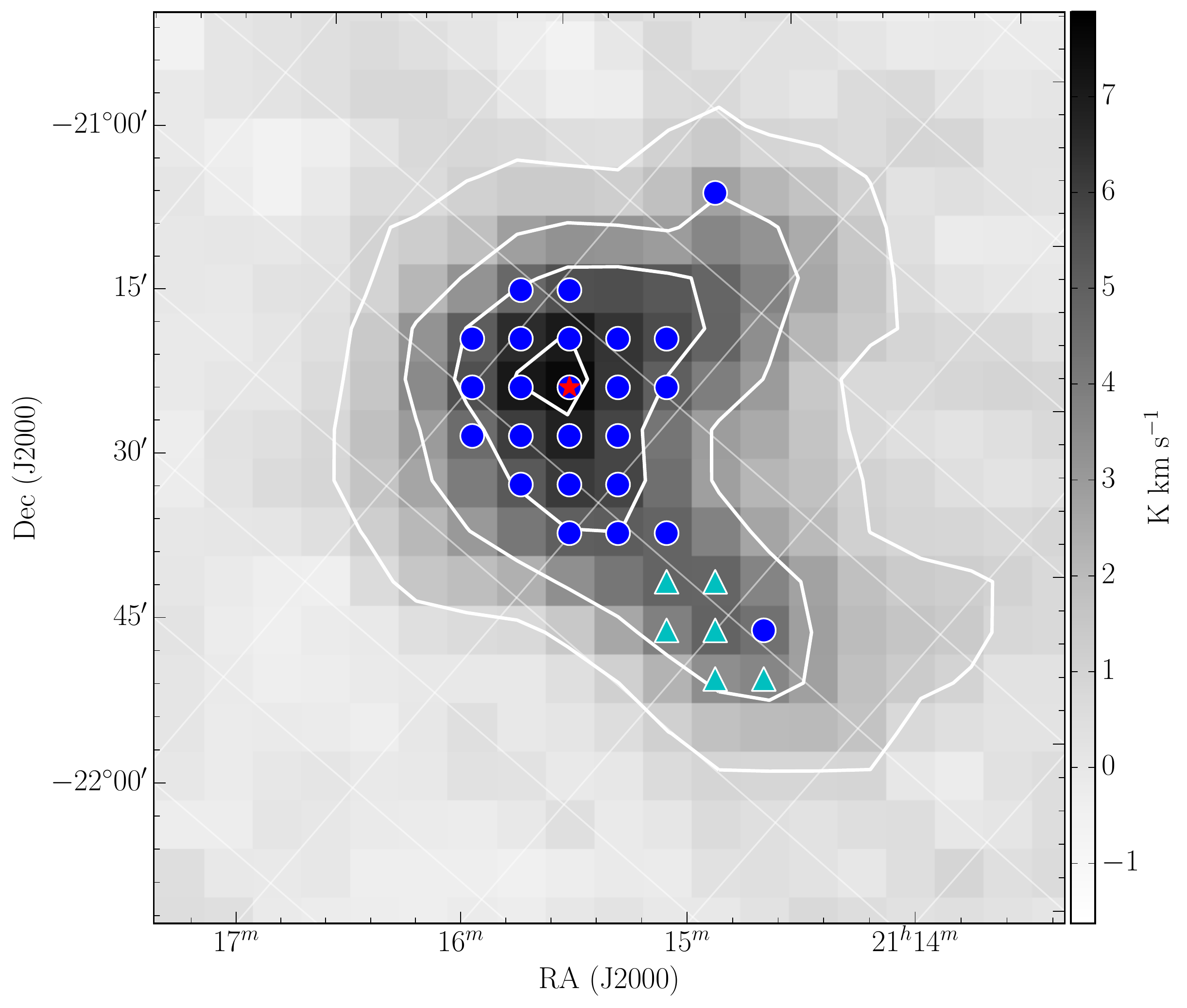}
\includegraphics[height=0.4\textwidth, angle=0, trim=0 0 0 0]{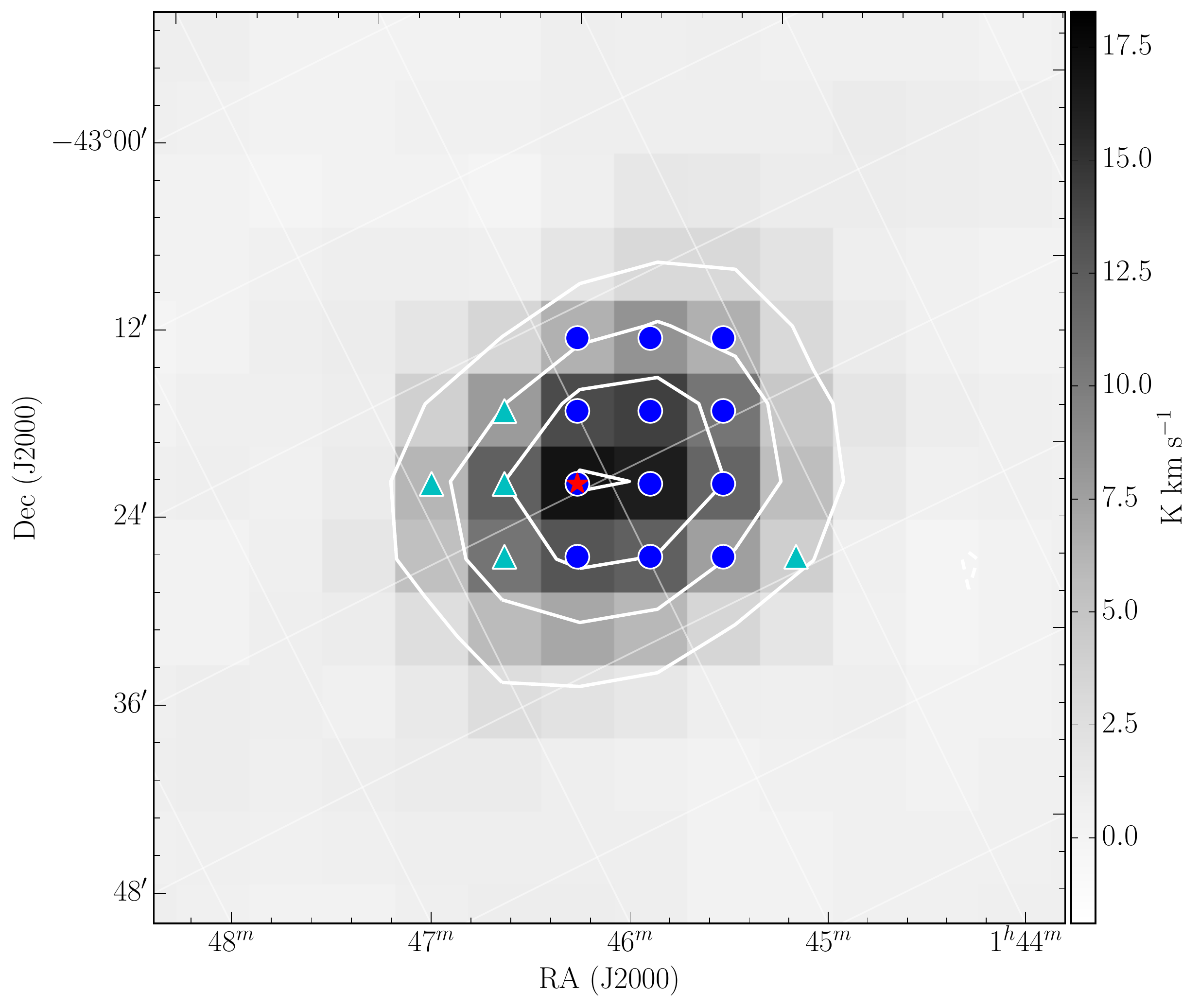}
\includegraphics[height=0.4\textwidth, angle=0, trim=0 0 0 0]{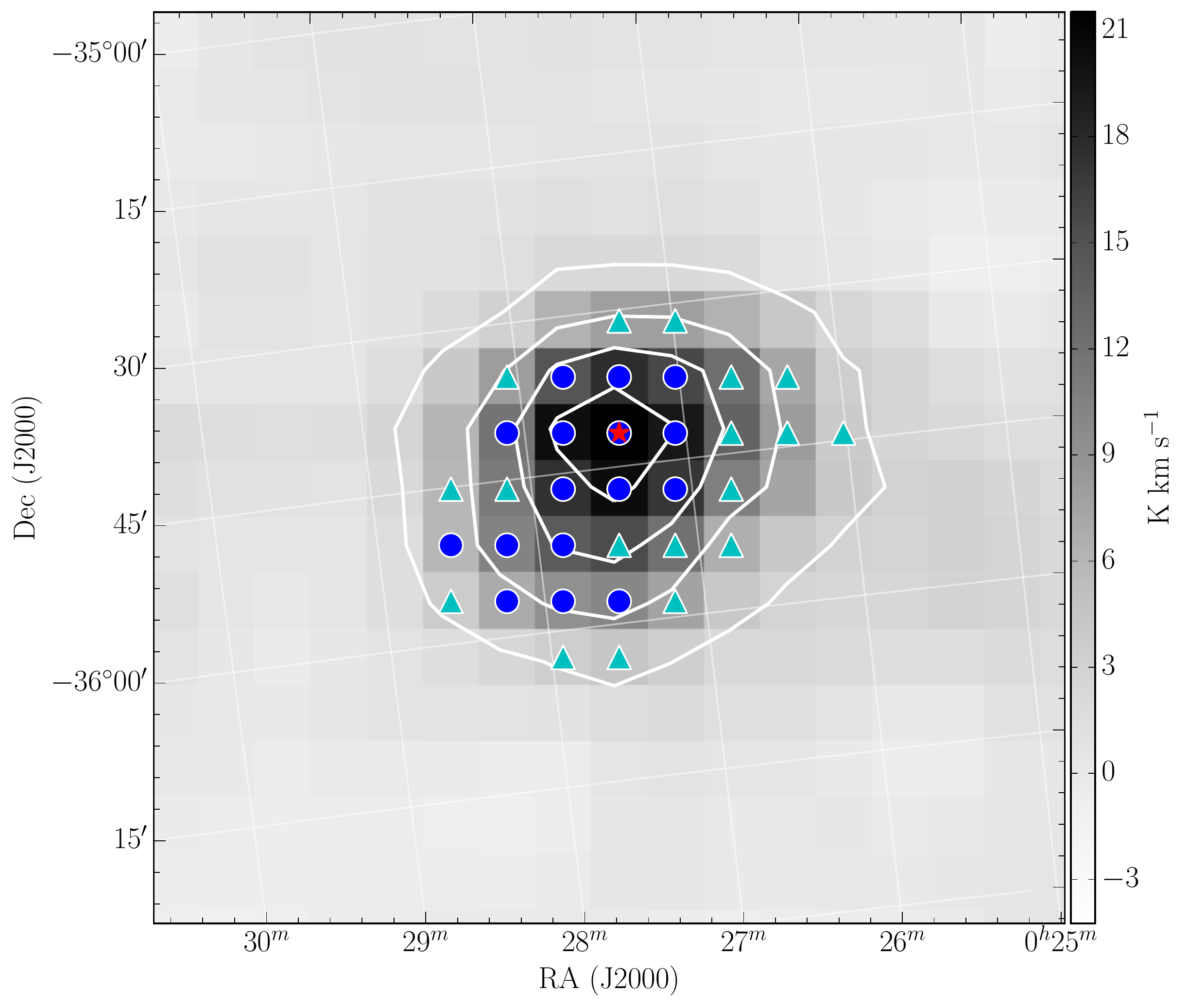}
\includegraphics[height=0.4\textwidth, angle=0, trim=0 0 0 0]{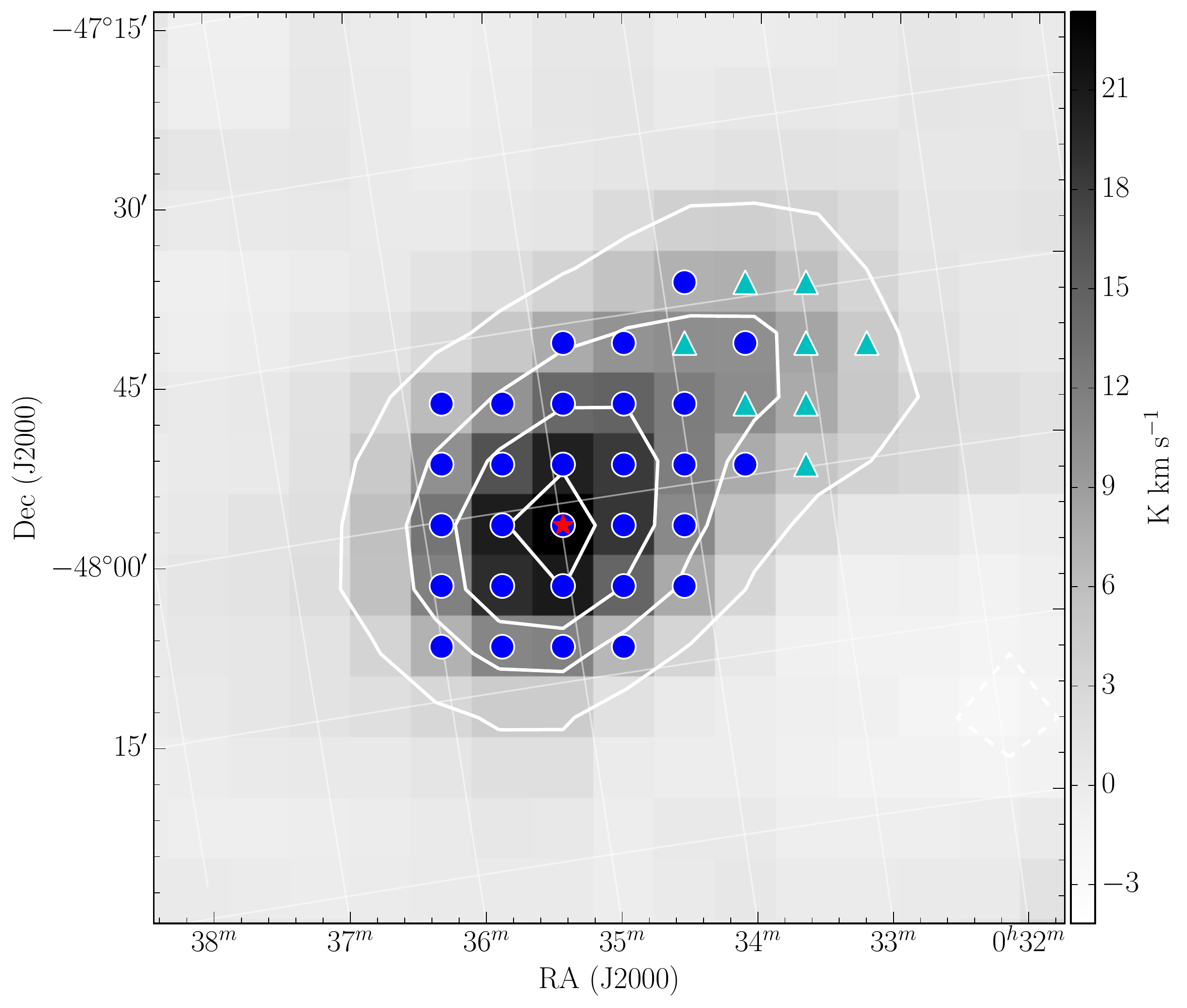}
\caption{\small Shown here are four examples of the clouds in the GASS HVC sample (ordered by increasing column density): GHVC~G028.1$-$41.1$-$221, GHVC~G273.0$-$70.2+375, GHVC~G332.2$-$80.0+119 and GHVC~G310.5$-$68.8+162. We have divided their spectral components into the two populations of dense and diffuse gas as described in Section \ref{fitting}, and find that components consistent with L02 generally lie on the outskirts and in the tails of GASS clouds, tracing where the denser HVC cores merge with a diffuse surrounding medium. The red star indicates the catalogued cloud core from \citet{Moss:2013bu}. The circles and triangles indicate the positions at which spectra were obtained of the cloud. Gas that is typical of the GASS HVC distribution in FWHM vs. $N_{HI}$ is shown as a blue circle, while gas that more closely follows L02 is shown as a cyan triangle. White contours at 4 equally-spaced levels from 25\% to the maximum of the image are shown to highlight the cloud.}
  \label{fig:gasscomponents}
\end{figure*}

To assess the overall comparability of the L02 data with GASS, we extracted GASS spectra at the position of all L02 spectra with declinations $\delta <$~0$^\circ$. We then fit a Gaussian to each GASS spectrum (both unsmoothed and smoothed to the same spatial and spectral resolution as L02) to obtain the fit parameters for comparison with L02. We excluded spectra that either had L02 velocities beyond the GASS range of $|v_{LSR}| <$~468\,km~s$^{-1}$ or with fitted peak brightness temperatures $T_{B,fit}$ $<$~171\,mK (3$\sigma$ in GASS) in L02 or in smoothed GASS data. This amounts to a total of 10 comparable spectra out of 83 spectra at southern declinations.  The results are shown in Figure \ref{fig:lockgass}, for $T_B$, $v_{LSR}$, FWHM and $N_{HI}$. 

Overall, there is good agreement between the two surveys with the majority of lines in GASS deviating by less than 20\% from the L02 measurement. The closest agreement is in velocity, confirming that each cloud has been detected in both surveys at the same velocity. There is some scatter in brightness temperature and line-width, also evident in the resulting column density; this largely reflects the fact that the Parkes beam is 50\% smaller than the 140\,ft beam and is mostly accounted for by the data smoothed to 21$'$ resolution. The mean GASS/L02 ratios for each property using the smoothed-GASS fit parameters are 1.09 ($T_B$), 0.97 (FWHM), 1.06 ($N_{HI}$) and 1.00 ($v_{LSR}$), indicating that there is a systematically brighter $T_B$ and slightly narrower FWHM measured in GASS compared with L02 (related to the above-mentioned beam dilution). \citet{Kalberla:2015kg} reported for their recently-reprocessed GASS~III a scaling factor of 0.96 in $T_B$ in comparison with GASS~II, which would correspondingly bring the $T_B$ ratio to $\sim$1.05 thus accounting for part of the deviation. However, given that the deviation is small and typically within the errors of GASS measurements, we conclude that the agreement between the two surveys is strong enough to justify direct comparison. 

\subsection{Separating the two populations}\label{fitting}
Our following investigation of the cloud populations is framed by two key questions: 1) where is each type of component (dense or diffuse) spatially located when we look across each GASS HVC, and 2) is the distribution across GASS HVCs of each type of component random or ordered in some way. To answer these questions, we use the cores of GASS HVCs and the L02 distribution as our starting points to quantitatively separate the two populations of neutral hydrogen. Figure \ref{fig:gasshist} shows an alternative view of the evident populations (at a constant $N_{HI}$), where GASS HVC cores (red) and L02 clouds (cyan) are visibly distinct. Our goal here is to develop a method that allows us to reliably quantify this separation and gain insight into the properties of dense and diffuse gas. 

Here we describe a machine learning approach we have adopted in order to provide a separation method that is statistically rigorous and offers probabilistic classifications for individual components. An advantage of this machine learning approach is that it simultaneously classifies in three dimensions ($N_{HI}$, $T_B$ and FWHM).

To determine which machine learning algorithm to adopt, we used the {\sc scikit-learn} package in Python \citep{Pedregosa:2011tv}. To form our test data set, we used the GASS cores and L02 data to represent the two distinct populations visible in Figure \ref{fig:allcomps}. We tested various classifiers both in $T_B$ vs. $N_{HI}$ and FWHM vs. $N_{HI}$ to check and visualise performance, using 60\% of the data to train and 40\% to test. In all cases, we obtained high scores of $\sim$90\% reclassification accuracy when applying the fit to the test sets independent of algorithm, with slight variation of a few percent maximum depending on the random division between training and test data. We adopted linear discriminant analysis (LDA) as the method for our distinction of populations, based on its relative simplicity and consistently high scores. LDA is similar to principal component analysis, but differs in that it is supervised (incorporating a training set) and finds axes that maximise the separation between the classes rather than the variance. In 2D with two classes, it determines a line separating the populations whereas in 3D it corresponds to a plane.

In order to apply this method and separate the populations most accurately, we tested on the 2D space of $f$($T_B$, $N_{HI}$) and $f$(FWHM, $N_{HI}$), as well as expanding the number of classifier features to include all three properties as $f$($T_B$, FWHM, $N_{HI}$). We again used the GASS cores and L02 points to train the classifier assuming these are dominated by the dense and diffuse populations respectively. To average out the effects of sampling and ensure a convergent fit, we ran the LDA classifier 10$^4$ times and took the average LDA fit properties. We then used this converged classifier to predict the classification of the GASS cores, the L02 data and the entire GASS data sample. This allowed us to check the performance of the classifier on both the known subsets and classify the wider sample in the different parameter spaces available. The resulting spectral classifications are given in Table \ref{tb:mlcompare}, with the 2D classification results visualised in Figure \ref{fig:ml2d}. Clearly indicated in this table is the fact that regardless of which parameter space we classify in, the spectral ratios are extremely consistent. Based on these results, we adopt the machine learning approach for $f$($T_B$, FWHM, $N_{HI}$) to make use of the full information available for classification. We give the specific {\sc scikit-learn} parameters of this 3D classifier in Table \ref{tb:lda}.

\begin{table*}
\centering
{
\caption{\small Resulting classifications from machine learning in the three cases: $f$($T_B$, $N_{HI}$), $f$(FWHM, $N_{HI}$) and $f$($T_B$, FWHM, $N_{HI}$). Regardless of the space in which we divide, we see strong consistency in the ratios. Shown are the number of components classified as each type (dense/diffuse) for the different subsamples, with the relative fraction given in brackets. In order to make full use of the information available, we adopt the $f$($T_B$, FWHM, $N_{HI}$) results as most representative of the sample.}\label{tb:mlcompare}
~\\
\begin{tabular}{lcccccccc}
\hline
 & Cores$_{dens}$ 	&  Cores$_{diff}$  &  L02$_{dens}$ & L02$_{diff}$ &  GASS$_{dens}$ & GASS$_{diff}$ \\ 
\hline
$f$($T_B$, $N_{HI}$)  				& 526 (0.99)		& 	8 (0.01)	&	 75 (0.25)	&	230 (0.75) & 9360 (0.90) & 1095 (0.10) \\
$f$(FWHM, $N_{HI}$) 					& 526 (0.99)		& 	8 (0.01)	&	 75 (0.25)	&	230 (0.75) & 9354 (0.89) & 1101 (0.11) \\
$f$($T_B$, FWHM, $N_{HI}$)		& 525 (0.98)		& 	9 (0.02)	&	 78 (0.26)	&	227 (0.74) & 9415 (0.90) & 1040 (0.10) \\

\hline
\end{tabular}}
\end{table*}

Using the LDA analysis for $f$($T_B$, FWHM, $N_{HI}$) on our total sample of 10455 usable GASS spectra, 9415 spectra are classified as part of the dense population and 1040 spectra are classified as part of the diffuse population, giving relative percentages of 90\% and 10\% respectively. A total of 145 GASS clouds ($\sim$27\%) have at least 1 diffuse spectral component, while only 8 clouds ($\sim$1\%) feature solely diffuse components. On average, diffuse components make up 28\% of the clouds in which they are found. The low number of isolated diffuse clouds is likely because GASS in general is not sensitive to diffuse gas components, which are by definition fainter than dense components. However, it is also likely that the diffuse gas tends to exist on the edges of clouds as supported by the findings of \citet{Murphy:1995hn}, where classical HVCs are the bright parts of a large-scale mostly diffuse halo medium that is not easily characterisable as isolated cloud structures.

\begin{figure}[t]
  \centering
  \includegraphics[width=0.48\textwidth, angle=0, trim=0 0 0 0]{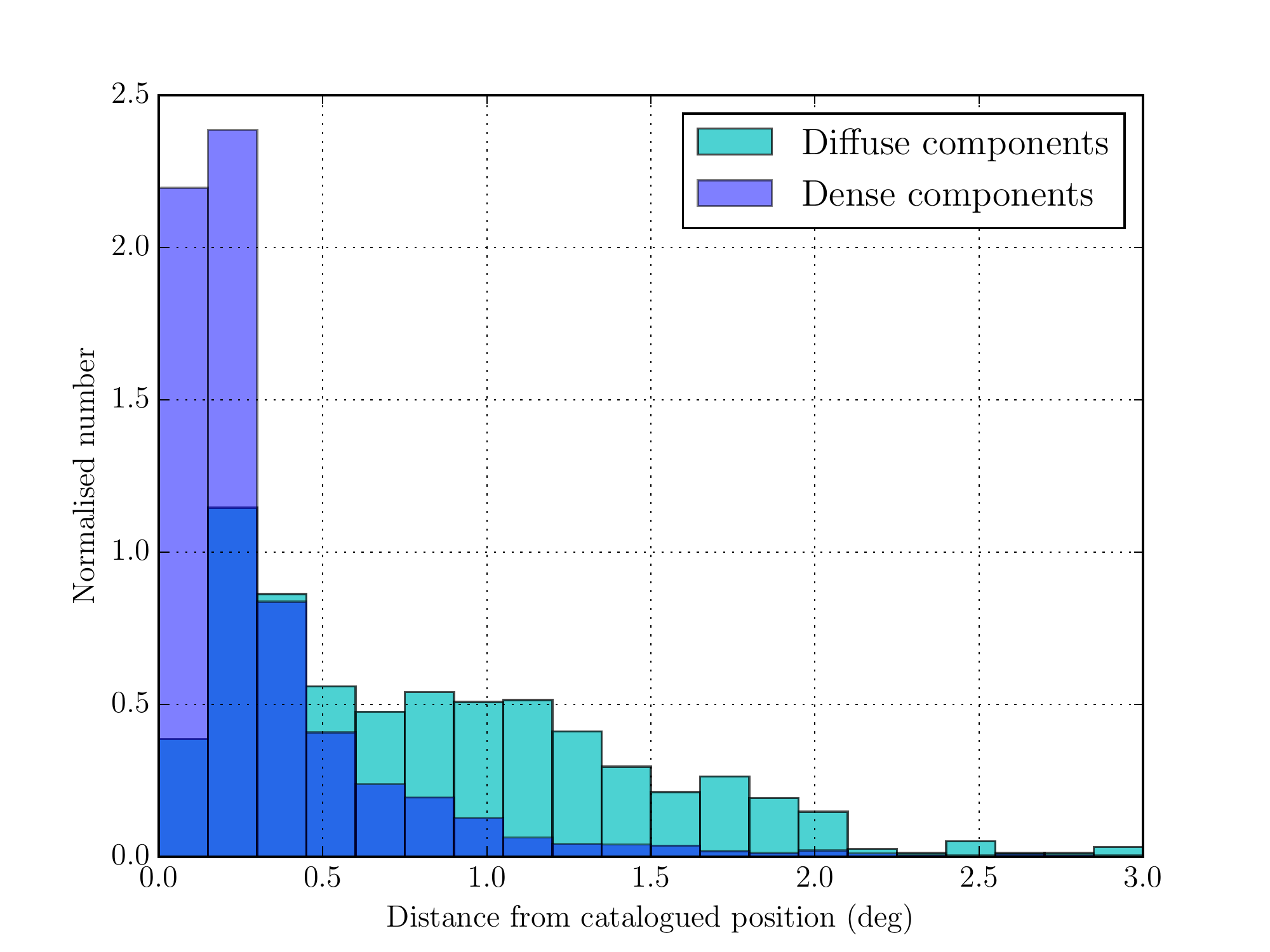}
  \caption{\small Normalised histogram of the angular distance between diffuse/dense spectral components and the cloud core of the GASS HVC with which the components are associated. There is a clear difference in the diffuse and dense populations of gas, where diffuse components are likely to be found around three times as far from the core of the HVC as the dense components. The median distance for dense gas is $\sim$0.2$^\circ$ and the median distance for diffuse gas is $\sim$0.7$^\circ$, while the  mean distance for dense gas is $\sim$0.3$^\circ$ and the mean distance for diffuse gas is $\sim$0.8$^\circ$. Thus at a given value of $N_{HI}$, the diffuse gas is found at the edges and tails of the HVCs while the dense gas is closer to the core.}
  \label{fig:distance}
\end{figure}

To examine spatial trends across clouds, we visualised the location of each type of gas on individual clouds and confirmed that in general we see diffuse components at the edges. We show four examples in Figure \ref{fig:gasscomponents}, spanning across the range in cloud column density. The blue circles represent the spectral components that correspond to the dense population, whereas the cyan triangles represent the diffuse components. We also consider the spatial separation of each population, by calculating the angular distance between the catalogued cloud core of each GASS HVC \citep{Moss:2013bu} and its associated spectral components. We show the results for this in Figure \ref{fig:distance}, with a normalised histogram for both the diffuse and dense populations. Quantitatively, we find that the median distance from the core for dense gas is $\sim$0.2$^\circ$ and the median distance for diffuse gas is $\sim$0.7$^\circ$, while the  mean distance for dense gas is $\sim$0.3$^\circ$ and the mean distance for diffuse gas is $\sim$0.8$^\circ$. In general we find that the diffuse component is displaced from the dense components (and thus from the core of the HVC) by $\sim$0.5$^\circ$.

\begin{table}[t]
\centering
{\footnotesize
\caption{Average LDA classification parameters from Python {\sc scikit-learn} for $f$($T_B$, FWHM, $N_{HI}$) based on 10$^4$ iterations of the training process. We adopted these values for the final classification on the entire GASS data set, as well as the HVC cores and L02 data.}\label{tb:lda}~\\
\begin{tabular}{lcccc}
\hline
 	&  Average &  $\sigma$ \\ 
\hline
coef\_  	& $[-1.03, +1.98, -0.91]$ 	 &  $[0.53, 0.52, 0.25]$\\
xbar\_  	&	$[-0.54, -0.66, -0.28]$ 	 &  $[0.04, 0.05, 0.04]$\\
scalings\_  	&	$[-0.47, +0.91, -0.42]$ 	 &  $[0.24, 0.23, 0.11]$\\
intercept\_  	&	2.22 & 0.19\\
\hline
\end{tabular}}
\end{table}

\subsection{Average spectrum of the diffuse and dense components}
Based on our analysis so far, we have separated all available GASS HVC spectral components into one of two populations: diffuse faint components that are similar to those found by L02, and dense components that are brighter, narrower and generally coincident with the cores of HVCs. To gain an insight into the physical properties of each population on a large scale, we examine their average spectrum. To do this, we obtained the actual spectrum corresponding to each GASS spectral component, shifted the peak fitted velocity to be centralised at 0\,km~s$^{-1}$, and over-plotted each spectrum onto a plot of either diffuse or dense components respectively. We only considered spectra which met the same criteria as our scatter plots: no association with flagged clouds, $|v_{LSR}| \ge$ 90\,km~s$^{-1}$ and $N_{HI} >$ 10$^{18}$\,cm$^{-2}$. In addition, there had to be complete spectral coverage of velocities $\pm$55\,km~s$^{-1}$ (diffuse) or $\pm$30\,km~s$^{-1}$ (dense) of the central peak (removing a total of 5 spectra). Due to contamination of bright signal evident through inspection (generally associated with low-velocity Galactic structure in the cubes of catalogued clouds rather than the clouds themselves), a total of 33 diffuse spectra with peak $T_B >$~1\,K were removed. With these filters applied, we are left with a total of 1003 diffuse components and 9414 dense components of a reduced total of 10417 spectra.

\begin{figure}[t]
  \includegraphics[width=0.48\textwidth, angle=0, trim=0 0 0 0]{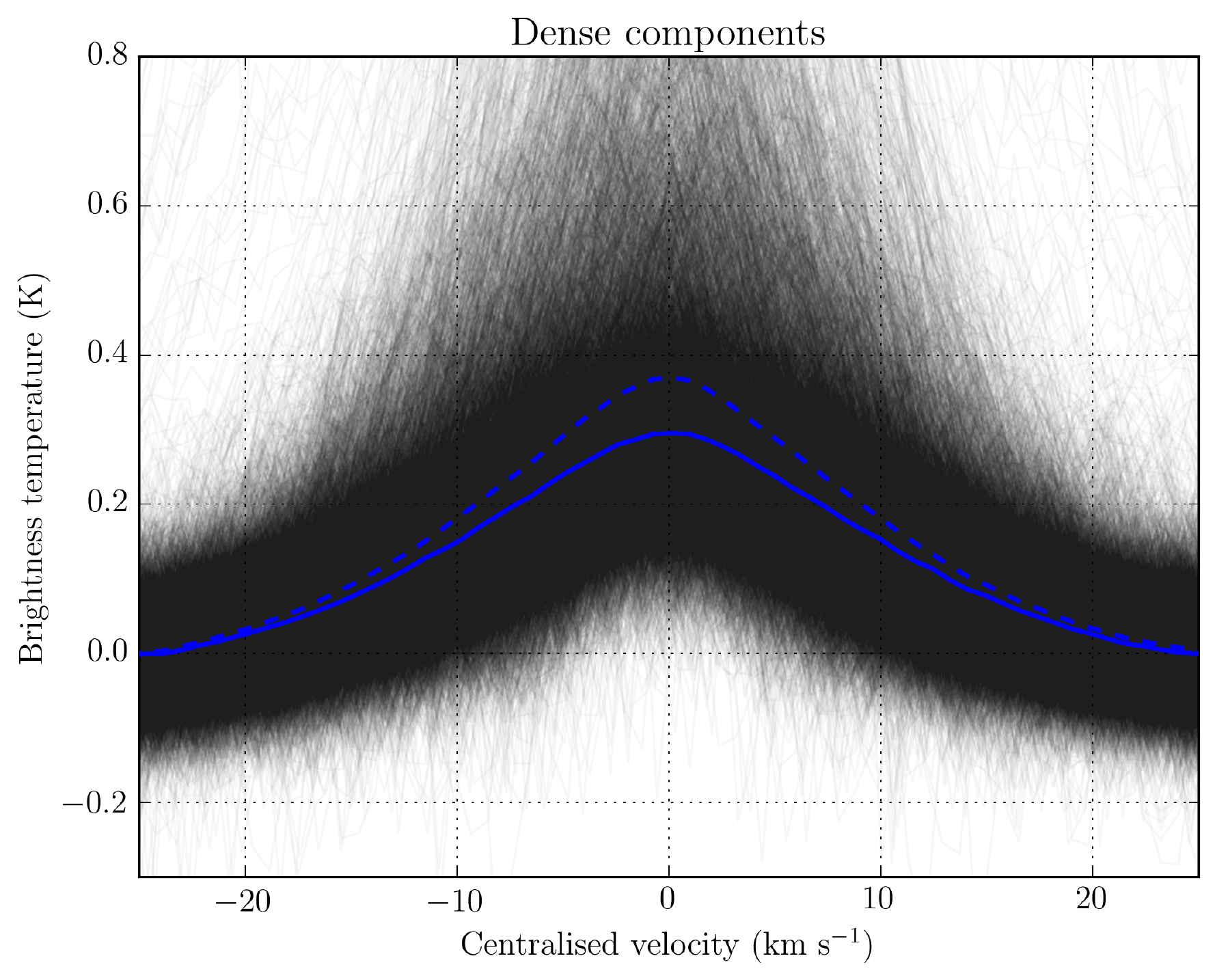}
  \includegraphics[width=0.48\textwidth, angle=0, trim=0 0 0 0]{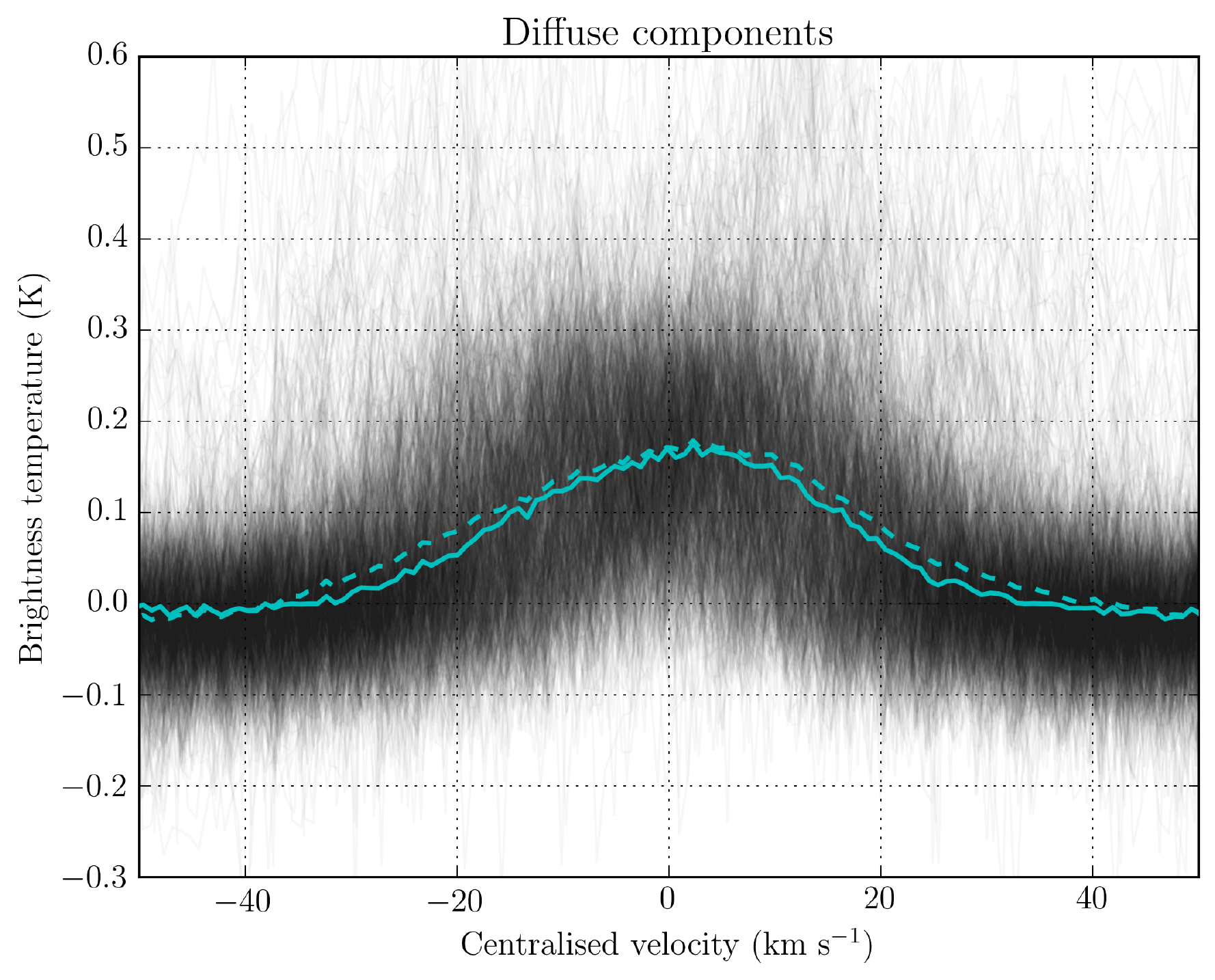}
  \caption{\small All spectral components incorporated into our analysis across the GASS HVCs are plotted in black, separated into either the dense (top) or the diffuse (bottom) population. We over-plot the mean spectrum (solid line) and median spectrum (dashed line) in each case, with the diffuse spectra in cyan and the dense spectra in blue. We see that the mean is skewed upwards by bright spectra (particularly affecting the dense population), and that the median spectrum is more visually representative of the data. The diffuse components show a large degree of scatter in their structure, whereas the dense components are generally clearer and more isolated. Note that the axes are on different scales for clarity of the spectral structure.}
  \label{fig:allspec}
\end{figure}

\begin{figure}[t]
  \includegraphics[width=0.48\textwidth, angle=0, trim=0 0 0 0]{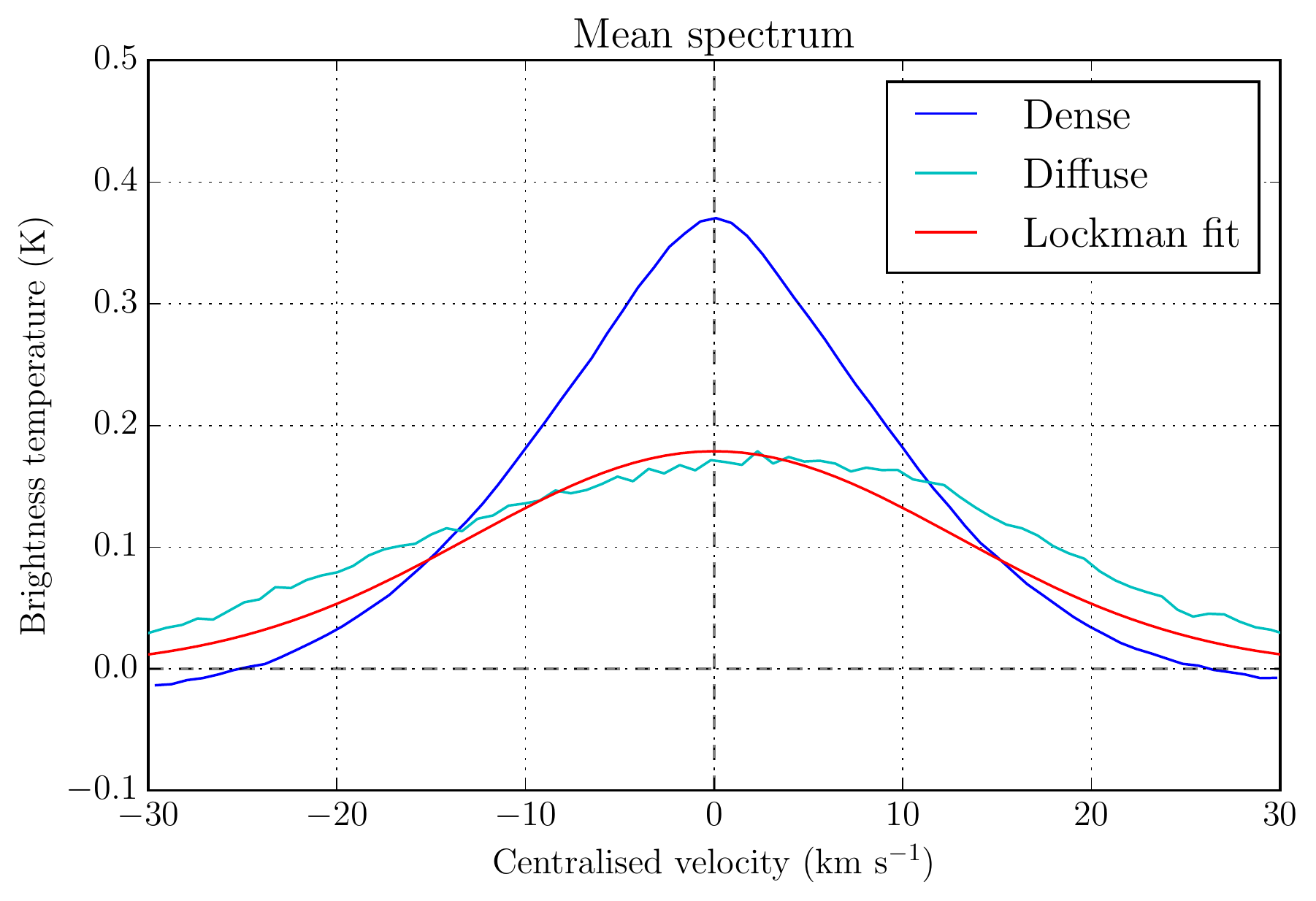}
    \includegraphics[width=0.48\textwidth, angle=0, trim=0 0 0 0]{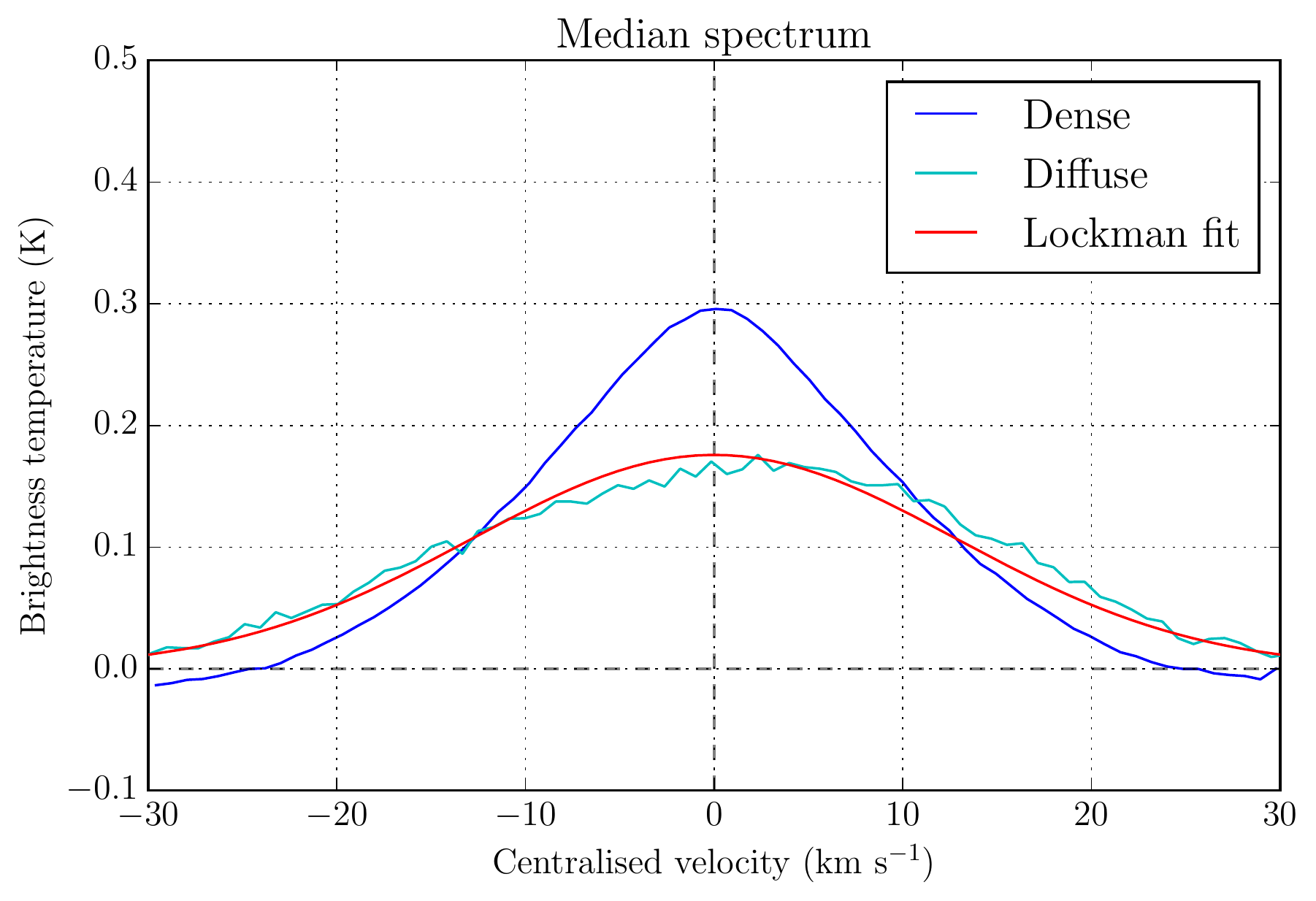}
  \caption{\small Simplified version of Figure \ref{fig:allspec} without the raw spectra shown. Shown are the mean spectra (top) and median spectra (bottom), for both dense components (blue) and diffuse components (cyan). Our determination of these populations means that by definition the diffuse spectrum is both broader and fainter, however here we see the clear quantitative difference between the populations. For reference, we also over-plot a spectrum representative of the properties of L02 (red solid line), scaled for comparison to the peak value of the mean/median diffuse spectrum (corresponding to a scaling factor of 3).}
  \label{fig:allspec2}
\end{figure}

We show the average (solid line) and median (dashed line) spectrum for both the diffuse (in cyan) and dense (in blue) population in Figure \ref{fig:allspec}, over-plotted on all spectra used to calculate both the average and the median spectrum. A simplified version without all spectra plotted is given in Figure \ref{fig:allspec2}, where the red line shows the average L02 spectrum scaled for comparison to the peak value of the mean/median diffuse spectrum (corresponding to a scaling factor of 3). The median spectrum is visually more representative of the bulk of the spectra, since it is less subject to skew by extremely bright spectra. Not surprisingly, we see that the diffuse components are on average fainter and broader and the dense components are brighter and narrower. 

\begin{table*}
\centering
{\footnotesize
\caption{Parameters of the average dense and diffuse GASS spectra as separated by the LDA method.}\label{tb:avprop}
\begin{tabular}{lcccc}
\hline
 & FWHM$_{diff}$ (km~s$^{-1}$) 	&  $T_{B,diff}$ (K) &  FWHM$_{dens}$ (km~s$^{-1}$) & $T_{B,dens}$ (K) \\ 
\hline
Median  	&	32		& 	0.17		&	 21		&	0.30\\
Mean 		&	37		& 	0.18		&	 21		&	0.35\\

\hline
\end{tabular}}
\end{table*}

We present the fit parameters to the median and mean GASS spectra in Table \ref{tb:avprop}. The fitted median/mean properties for the diffuse gas differ most due to the higher standard deviation in the population. Our results in both cases show that the average dense spectrum is close to the standard spectrum of known HVCs in other surveys of $\sim$20--25\,km~s$^{-1}$ \citep{Wakker:1997ha,Kalberla:2006kd,BenBekhti:2006ef,Moss:2013bu}, while the diffuse components are probing neutral hydrogen in a significantly different state that is not typically sampled by surveys of HVC gas. The diffuse spectrum is broader and fainter, and the spatial location in GASS appears to trace the transition between the cores of HVCs and where they merge with the surrounding medium. We also see that the L02 population (with a median FWHM of 30.3\,km~s$^{-1}$ and a median $T_B$ of 57\,mK) is clearly similar to the diffuse spectrum (but at a fainter level due to the lower sensitivity of GASS to the diffuse population), with the neutral hydrogen likely to be in the same physical state as our GASS diffuse components.

It is worth noting here (as also mentioned above) that the angular resolution of a survey makes a difference to the measured FWHM. Surveys at $>$ 30$'$ resolution find, in general, broader spectra due to the merging at this angular scale of the diffuse and dense spectra. This has been pointed out by \citet{Wakker:1997ha}, and was also found by \citet{Moss:2013bu} in addressing incongruities in the spectral distribution between high-velocity gas found in GASS with a median FWHM of $\sim$19\,km~s$^{-1}$ and the Leiden/Argentine/Bonn (LAB) survey with a median FWHM of $\sim$28\,km~s$^{-1}$ \citep{Kalberla:2006kd}. Given our result here where many HVCs can be separated into core components surrounded by diffuse components and the average HVC size at GASS sensitivity of $\sim$1~deg$^2$, the difference in measured line-width as a function of angular resolution is thus understandable.

While the observed H{\sc i} line is in general a combination of both thermal and turbulent broadening, thermal broadening alone cannot explain the diffuse spectrum. To show this, we first convert from FWHM line-width to velocity dispersion $\sigma_\nu$ by dividing by $2 \sqrt{2 \ln 2} \approx 2.35$, and then estimate the kinetic temperature using the equation

\begin{equation}
T_{\rm k} = \frac{m_{\rm H} \sigma_\nu^2}{k},
\end{equation}

where $m_{\rm H}$ is the mass of the hydrogen atom and $k$ is the Boltzmann constant. For the dense gas at a line width of 20\,km~s$^{-1}$, $T_{\rm k} \sim 9 \times 10^3$\,K, and the gas can be mainly neutral. However, for the diffuse gas with a line width of 30\,km~s$^{-1}$, $T_{\rm k} \sim 2 \times 10^4$\,K which is well above the temperature at which hydrogen is expected to become fully ionised and thus indicates that the diffuse gas H{\sc i} line is broadened through turbulence rather than temperature. This makes sense when combined with the knowledge that the diffuse gas tends to trace the edges of clouds which are both interacting with a hot halo medium and potentially being ram-pressure stripped due to their motion through this medium.

\begin{table*}[t]
\centering
{\caption{\footnotesize Mass fraction estimates for neutral hydrogen in the halo, assuming properties of the dense/diffuse H{\sc i} population as derived from GASS and/or L02. $N_{HI}$ represents the average column density of each population either calculated from GASS ($N_{HI,GASS}$) or L02 ($N_{HI,L02}$). $f$ represents the fraction of dense to diffuse gas, similarly derived from GASS (9:1) or L02 (1:3). $\xi$ is the resulting mass fraction obtained by multiplying these two properties together and normalising the result. We have highlighted in grey the most reliable mass fraction estimate based on the assumption that L02 better probes the true ratio of dense to diffuse gas and GASS provides a better statistical measure of the column densities associated with each population.}\label{tb:massest}
~\\
\begin{tabular}{lcccccccc}
\hline
 &	$N_{HI,dens}$ (cm$^{-2}$) & $N_{HI,diff}$ (cm$^{-2}$) & $f_{dens}$ & $f_{diff}$ 	&  $\xi(N_{H,dens})$ & $\xi(N_{H,diff})$\\ 
\hline
$N_{HI,L02}~|~f_{GASS}$ 	&  [1.92--2.73] $\times$ 10$^{19}$ & [0.23--0.50] $\times$ 10$^{19}$ &	0.90 		&  	0.10		&	 0.97--0.99		&	0.01--0.03\\
$N_{HI,L02}~|~f_{L02}$ 		& [1.92--2.73] $\times$ 10$^{19}$ & [0.23--0.50] $\times$ 10$^{19}$&	0.26 	& 		0.74		&	 0.57--0.81		&	0.19--0.43\\
$N_{HI,GASS}~|~f_{GASS}$ 	& [1.22--1.43] $\times$ 10$^{19}$ &	[1.06--1.29] $\times$ 10$^{19}$ & 0.90 		&  	0.10		&	 0.89--0.92		&	0.08--0.11\\
\rowcolor{Gray}
$N_{HI,GASS}~|~f_{L02}$  & [1.22--1.43] $\times$ 10$^{19}$& [1.06--1.29] $\times$ 10$^{19}$	&	0.26			& 		0.74		&	 {0.25--0.32}		&	{0.68--0.75}\\

\hline
\end{tabular}}
\end{table*}

\subsection{Sky-covering fraction of the diffuse and dense halo gas}
\begin{figure*}
  \centering
  \includegraphics[width=0.95\textwidth, angle=0, trim=0 0 0 0]{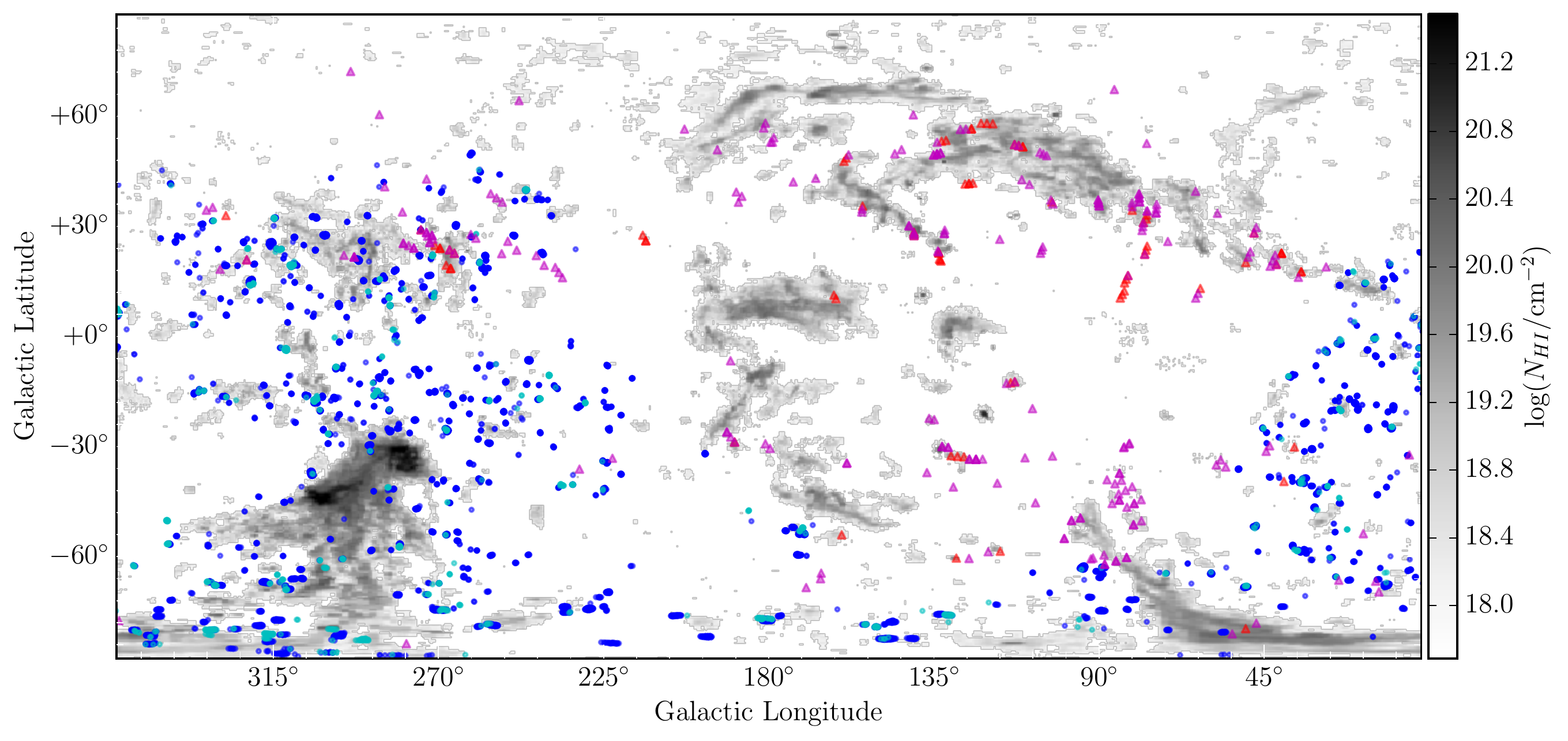}
  \caption{\small The distribution of both diffuse and dense spectral components on the sky. All components incorporated from both surveys are plotted, separated into the categories of GASS dense (blue circles), GASS diffuse (cyan circles), L02 dense (red triangles) and L02 diffuse (magenta triangles). Both types of gas are found all over the sky, with no obvious region of preference. The background map is the integrated-intensity Leiden/Argentina/Bonn survey of H{\scshape i} \citep{Kalberla:2005de}, excluding emission at deviation velocities $| v_{dev} | <$ 75\,km~s$^{-1}$. This map was produced and provided by Tobias Westmeier, from the methods described in \citet{Westmeier:2007ve}.}
  \label{fig:distsky}
\end{figure*}

We now consider the distribution of dense and diffuse gas on the overall Galactic H{\scshape i} sky. We plot both samples in Figure \ref{fig:distsky}, with the GASS population shown in blue (dense components) and cyan (diffuse components) circles and L02 shown in red (dense components) and magenta (diffuse components) triangles. Both dense and diffuse H{\sc i} are found all over the high latitude sky, with no obvious region of preference. Of key interest with regard to the question of neutral gas available for Galactic accretion is to determine the relative mass fraction of dense or diffuse H{\sc i}. To estimate this mass fraction (represented here as $\xi$), we adopt the following equation:

\begin{equation}
\xi = f N_{HI},
\end{equation}

where, for dense or diffuse gas, $f$ is the fraction of components measured from our data and $N_{HI}$ is the measured average column density for that population. The main purpose of this section is to determine the best values for $f$ and $N_{HI}$ using our joint analysis of the GASS and L02 data.

In the GASS HVCs, we found that separation of the sample based on the properties of lines in L02 results in 90\% GASS HVC spectra classified as dense components, while 10\% are classified as diffuse components that overlap with the faint population revealed in L02. Thus we can conclude, using our LDA separation method, that GASS contains dense/diffuse proportions of roughly 9:1. We see that the dense components dominate the GASS sample, which is a strong function of the limited sensitivity of GASS to faint diffuse components compared with L02. Conversely, the majority of the sensitive L02 sample falls into the diffuse category. However, since the survey randomly sampled the halo, we do expect some fraction of sightlines through dense neutral hydrogen. To put a number on this, we count all L02 spectra using the machine learning approach in Section \ref{fitting} that are classified as dense or diffuse. Of these 305 components, we find that 78 are likely associated with the dense population and 227 are likely associated with the diffuse population.  If we convert these to percentages, than we see that the fraction is reversed compared to the GASS sample, with 26\% dense components and 74\% diffuse components. 

The more sensitive L02 indicates (based on a random sample through gas in the halo) that diffuse gas dominates over dense gas in the halo with a ratio of 3:1, suggesting that every dense component is outweighed by a diffuse medium that covers 3 times as much area on the sky. In GASS with reduced sensitivity to diffuse gas, we find the opposite ratio, where dense to diffuse is in the ratio of 9:1. Thus in general, due to sensitivity limits and selection effects, we could be sampling a very small fraction of high velocity H{\scshape i} based on the GASS HVCs alone. Considering the result of \citet{Lehner:2012fc}, who found that 74\% of their HVC directions featured column densities $N_{HI} <$ 3 $\times$ 10$^{18}$\,cm$^{-2}$, we would similarly expect that GASS would be sensitive to $\lesssim$ 25\% of the total H{\scshape i} in the halo. Regardless of the exact fraction, it is clear here that studies of HVCs based on large-scale relatively shallow surveys will not accurately sample the diffuse gas, which explains why UV absorption studies through the Galactic halo \citep[e.g.][]{Savage:1996dz,Wakker:2003fx,2006ApJS..165..229F,Collins:2009co,Lehner:2011ic} have found evidence for the presence of more H{\sc i} than suggested by typical surveys of H{\sc i} high-velocity gas in the halo.

We can estimate the median/mean column density of diffuse and dense gas from the averaged GASS spectral components, assuming their large number to be representative of the two populations. These are $N_{H,dens}$ = [1.22, 1.43] $\times$ 10$^{19}$\,cm$^{-2}$ and $N_{H,diff}$ = [1.06, 1.29] $\times$ 10$^{19}$\,cm$^{-2}$, respectively. We estimate the mass fraction $\xi$ from the product of the column density and H{\sc i} fraction (derived from the ratio of dense to diffuse). The similar column density for both dense and diffuse gas implies that, given the larger sky-covering factor of diffuse gas, the total mass associated with the diffuse H{\scshape i} is likely to be 2--3 times that of the total dense H{\scshape i} (under the assumption that the more sensitive L02 is better representative of the ratio of dense to diffuse gas). Alternatively but far less likely, dense H{\sc i} may significantly outweigh the diffuse H{\sc i} and dominate the halo if we assume that GASS better traces both the column density and the ratio. The average column densities based on L02 are $N_{H,dens}$ = [1.92, 2.88] $\times$ 10$^{19}$\,cm$^{-2}$ and $N_{H,diff}$ = [0.23, 0.50] $\times$ 10$^{19}$\,cm$^{-2}$ for the median and mean properties respectively. Repeating the same analysis as above, we find that the diffuse gas may be a tiny fraction of the total H{\scshape i} if the GASS spectral components represent the true ratio of dense and diffuse gas, with a more extreme divide due to an overall higher column density of dense gas as measured in L02. However, if we assume that L02 is better representative of the dense and diffuse proportions in the halo as well as their column densities, then we find the mass of dense gas to be around 2--3 times that of the diffuse gas.

We have summarised these mass fraction estimates in Table \ref{tb:massest}, assuming that: 1) either GASS or L02 better represent the relative fraction $f$ of dense and diffuse gas, and 2) either the average $N_{HI}$ of GASS or L02 is better representative of the overall population. The ranges are given by using the mean or median spectral properties. The range in mass fraction $\xi$ is based on assuming either value for the column density and is used as an indication of the uncertainty in our result. 

Overall we propose that L02 with its higher sensitivity probes closer to the relative proportions of dense versus diffuse gas in the halo of the Milky Way, while GASS with its large number of spectra offers a better estimation of the average properties of these populations. We conclude that the diffuse gas contributes at least as much mass (estimated as the product of column density and ratio) as the dense gas, but more likely 2--3 times as much H{\scshape i}. There remains inherent uncertainty in the exact mass fraction due to the intrinsic scatter within each population and the difficulty in probing the diffuse population with GASS (or any relatively shallow all-sky survey), however the 1:3 proportion of dense to diffuse components in L02 as revealed in our analysis is a strong indicator that diffuse gas outweighs dense gas in the halo.

Our results also highlight the danger in using sensitivity-limited surveys to measure mass: the same analysis based on GASS alone would suggest that dense gas dominates the halo, which is unlikely to be true given what we have shown here. We hope to extend this work in future to replicate the L02 survey in the southern hemisphere using Parkes or the Australia Telescope Compact Array, as well as investigate the hydrogen probed by the Effelsberg-Bonn H{\sc i} Survey \citep{Winkel:2016bn}, a northern complement to the GASS data. The recently released all-sky HI4PI \citep{Collaboration:2016jn} which combines GASS and EBHIS would be an excellent data set in which to investigate our results regarding the halo and mass fraction further.


\section{Summary}\label{conclusiony}
We have combined two samples of neutral hydrogen in the halo of the Milky Way, in order to better understand the structure of halo gas: the very sensitive 21 centimetre Green Bank survey for high-velocity H{\scshape i} \citep{Lockman:2002bu}, and a catalogue of bright HVCs identified in the Galactic All Sky Survey \citep{Moss:2013bu}. These two surveys of H{\sc i} gas with their very different properties provide a complementary picture of the halo, with the L02 distribution sampling the halo with high sensitivity ($\sigma$ = 3.4\,mK) at random positions and the GASS HVCs sampling bright condensations of gas in the halo above a brightness of 4$\sigma$ from GASS ($\sigma$ = 57\,mK). As such, we gain a new insight into the distribution of both dense (bright, narrow line-widths and relatively cold) and diffuse (faint, broad line-widths and relatively warm) high-velocity neutral hydrogen gas in the Milky Way halo. 

We summarise the key results from this paper below:

\begin{enumerate}
\item We have discovered in the GASS sample evidence for two distinct populations of halo H{\sc i} which separate at a given column density: the dense distribution which is represented in most GASS spectra, and the diffuse distribution which aligns with L02.
\item In the GASS data, we find that dense components outnumber diffuse components in the ratio of 9:1. Conversely, in the L02 data, the dense to diffuse ratio is 1:3. This is reflective of the sensitivity limits of the GASS data.
\item We find that the low-brightness, broad-FWHM population which dominates L02 corresponds largely with the edges and tails of the GASS HVC population, where HVCs merge with the prevalent diffuse gas of the Galactic halo. The angular distance of diffuse components from the cloud core is on average 0.5$^\circ$ further than the dense gas.
\item We calculated the physical properties of the mean and median spectrum of both populations, finding that the dense distribution of gas is typical of classical HVCs, with a FWHM $\sim$ 20\,km~s$^{-1}$ and a brightness temperature of $T_B \sim$ 0.3\,K, while the diffuse distribution has a FWHM $\sim$ 30\,km~s$^{-1}$ with a brightness temperature of $T_B \sim$ 0.2\,K.
\item In determining the mass fraction of both dense and diffuse gas, we assume that L02 with its high sensitivity better traces the ratio of diffuse to dense gas while GASS with its large number of spectra better traces the average physical properties. Under those assumptions, we find that the mass fraction of diffuse H{\scshape i} is likely to be 2--3 times that of the dense H{\scshape i} in the halo.
\end{enumerate}

Based on these results for the Milky Way, we have confirmed observationally the existence of two populations of neutral gas in the Galactic halo. The dense distribution is typical of the majority of bright high-velocity gas, while the diffuse distribution is more widespread and far more difficult to detect, although often associated with the edges of the dense gas based on our results from GASS. In this picture, we can understand the classical HVCs as the bright dense cores within a ubiquitous diffuse medium surrounding most complexes of neutral hydrogen, tracing their interactions with the surrounding hot halo environment. These results are consistent with the findings of \citet{Murphy:1995hn}, as well as findings at other wavelengths of the presence of low-column density H{\sc i} not typically detected in large-scale H{\sc i} surveys. These results also agree well with previous findings arguing that classical HVCs cannot account for the Galactic star formation rate alone, with the largest HVC complexes contributing at most $\sim$0.4\,M$_\odot$~yr$^{-1}$ \citep{Putman:2012bsa}, compared with the estimated star formation rate of the Milky Way of $\sim$1\,M$_\odot$~yr$^{-1}$ \citep{Robitaille:2010hs}. 

The discovery of the prevalence of diffuse gas around HVCs, as we have shown here, takes us some of the way closer to reconciling this disparate star formation rate and gas infall rate. However the origin of the diffuse gas we detect, like the HVCs to which is it related, remains in question. Future multi-wavelength observations of the Galactic halo as well as the haloes of nearby galaxies will ultimately help resolve this question of origin and shed further light on the nature of the diffuse gas.


{\small \acknowledgments
We would like to thank our anonymous referee and statistics editor for their comments to help improve this paper. The Parkes Radio Telescope is part of the Australia Telescope National Facility which is funded by the Commonwealth of Australia for operation as a National Facility managed by CSIRO. The 140\,ft Telescope is operated by the National Radio Astronomy Observatory, which is a facility of the US National Science Foundation operated under cooperative agreement by Associated Universities, Inc. VAM acknowledges support from the University of Sydney, CSIRO Astronomy and Space Science and the ARC Centre of Excellence for All-Sky Astrophysics (CAASTRO) during the completion of this work. NMM acknowledges the support of the Australian Research Council through grant FT150100024. VAM thanks A.S.G. Robotham, B.J. Brewer and R. Lange for useful insights into machine learning that informed our adopted approach. The authors thank T. Westmeier for providing his map of the HVC sky seen in Figure \ref{fig:distsky}. This research made use of {\sc APLpy}, an open-source plotting package for Python hosted at http://aplpy.github.com, and the {\sc miriad} software package developed by the ATNF. We would also like to thank J.R. Dawson, A.S. Hill, J.R. Allison, G.A. Rees and E.K. Mahony for helpful discussions and feedback during this research.

\bibliographystyle{apj}

\end{document}